\documentclass[12pt]{article}

\usepackage{graphicx,amssymb,eucal}
\usepackage{url}

\def\nn{\nonumber}

\renewcommand{\thefootnote}{\fnsymbol{footnote}}

\newcommand{\tr}{{\rm Tr}}

\newcommand{\wt}{\widetilde}

\newcommand{\del}{\partial}
\newcommand{\ra}{\rightarrow}

\newcommand{\lra}{\leftrightarrow}
\newcommand{\half}{\frac{1}{2}}

\def\tr{\mathop{\rm tr}\nolimits}

\newcommand{\cA}{{\cal A}}

\newcommand{\cV}{{\cal V}}
\newcommand{\cD}{{\cal D}}

\newcommand{\bR}{\mathbb{R}}

\newcommand{\cO}{{\cal O}}

\newcommand{\bvo}{{\bar{v}_0}}
\newcommand{\X}{{X^1}}
\newcommand{\vnoc}{{\widetilde{v}_{0C}^n}}
\newcommand{\vmoc}{{\widetilde{v}_{0C}^m}}
\newcommand{\voc}{{\widetilde{v}_{0C}^1}}

\newcommand{\Bnmu}{{\bar{B}^n_\mu}}
\newcommand{\Bnnu}{{\bar{B}^n_\nu}}
\newcommand{\Bmmu}{{\bar{B}^m_\mu}}
\newcommand{\Bmnu}{{\bar{B}^m_\nu}}
\newcommand{\Blmu}{{\bar{B}^l_\mu}}
\newcommand{\Blnu}{{\bar{B}^l_\nu}}
\newcommand{\Bmo}{{\bar{B}^m_0}}
\newcommand{\Bno}{{\bar{B}^n_0}}

\newcommand{\Xm}{{X^m}}
\newcommand{\Xn}{{X^n}}

\newcommand{\psim}{{\psi_{m}}}
\newcommand{\psin}{{\psi_{n}}}
\newcommand{\psil}{{\psi_{l}}}
\newcommand{\dmu}{{\partial_\mu}}
\newcommand{\dnu}{{\partial_\nu}}
\newcommand{\dpsim}{{\dot{\psi}_m}}
\newcommand{\dpsin}{{\dot{\psi}_n}}

\newcommand{\dPsim}{{\dot{\Psi}_m}}
\newcommand{\dPsin}{{\dot{\Psi}_n}}
\newcommand{\phio}{{\phi_0}}
\newcommand{\Ukk}{U_{\rm KK}}

\newcommand{\be}{\begin{equation}}
\newcommand{\ee}{\end{equation}}
\newcommand{\bea}{\begin{eqnarray}}
\newcommand{\eea}{\end{eqnarray}}

\def\matt[#1,#2,#3,#4]{\left(%
\begin{array}{cc} #1 & #2 \\ #3 & #4 \end{array} \right)}

\def\v2#1{\vv2[#1]}
\def\vv2[#1,#2]{\left(%
{#1 \atop #2}\right)}

\begin{document}
\begin{titlepage}

\begin{flushright}
  {\tt hep-th/0608046}
\end{flushright}
\vspace{0.5in}

\begin{center}
  {\large \bf Dense Hadronic Matter in Holographic QCD} \\ \vspace{10mm}
  Keun-Young Kim$^a$, Sang-Jin Sin $^b$  and Ismail Zahed$^a$\\
\vspace{5mm}
$^a$ {\it Department of Physics and Astronomy, SUNY Stony-Brook, NY
11794}

$^b$ {\it Department of Physics, Hanyang University, Seoul 133-791,
Korea}\\

\vspace{10mm}

\end{center}

\begin{abstract}
We provide a method to study hadronic matter at finite density in
the context of the Sakai-Sugimoto model. We introduce the baryon
chemical potential through the external $U(1)_v$ in the induced
(DBI plus CS) action on the D8-probe-brane, where baryons are
skyrmions. Vector dominance is manifest at finite density. We
derive the baryon density effect on the  energy density, the
dispersion relations of pion and vector mesons at large $N_c$. The
energy density asymptotes to a constant at large density
suggesting that dense matter at large $N_c$ freezes, with the
pion velocity dropping to zero. Holographic dense matter enforces
exactly the tenets of vector dominance, and screens efficiently
vector mesons. At the freezing point the $\rho-\pi\pi$ coupling
vanishes with a finite rho mass of about 20\% its vacuum value.

\end{abstract}

\end{titlepage}

\renewcommand{\thefootnote}{\arabic{footnote}}
\setcounter{footnote}{0}


\newpage

\section{Introduction}
Recently there has been much interest~\cite{many,Sakai}
in the AdS/CFT approach~\cite{Maldacena} to study
the non-perturbative aspects of gauge theories at large
$N_c$ and strong coupling with even applications to heavy
ion collision experiments such as jet quenching~\cite{sz}. On the other hand,
dense QCD is of much relevance to compact stars and current heavy ion
collisions. This subject has been intensely studied in the past decade
following the observation that at asymptotic densities QCD matter may
turn to a color superconductor due to asymptotic freedom~\cite{BCS}. At very
large $N_c$ there is evidence that the Overhauser effect takes over with
the formation of density waves~\cite{OVER}.
Indeed, at weak coupling the high degeneracy
of the Fermi surface causes quark-quark (antiquark-antiquark) pairing of
the BCS kind with different color-flavor arrangements, while at large
number of colors quark-antiquark pairing of the Overhauser kind is favored.
In strong coupling, first principle calculations are elusive and the two
pairings may compete.

To address the issue of strongly coupled QCD at large $N_c$ in dense
matter we explore in this paper the holographic principle. In recent
years, the gauge/string duality has provided a framework for addressing
a number of problems in strong coupling where little is known from
first principles in the continuum. To do so, we consider a model for
holographic QCD recently discussed by Sakai and Sugimoto. In the limit
$N_f\ll N_c$, chiral symmetry in QCD is generated by immersing
$N_f$ D8-$\overline{\rm D8}$ into a D4 background in 5 dimensions
where supersymmetry is broken by compactification (Kaluza-Klein
mechanism with $M_{KK}$ scale). The induced DBI action on D8 yields
a 4 dimensional effective theory of pions and infinitly many vector
mesons where the $M_{KK}$ scale plays the role of an upper cutoff
(the analogue of the chiral scale $4\pi\,f_{\pi}$). At large $N_c$
baryons are Skyrmions.

In this paper, we extend the analysis by
Sakai and Sugimoto~\cite{Sakai} to finite baryon density.
Although there have been many  papers~\cite{chemical} on finite
chemical potential in the holographic approach, all involve the
R-charge or isospin instead of fermion number.
The difficulty for the fermion number is that
the $U(1)$ charge for R-symmetry is already dual to the
$U(1)$ charge of AdS Reisner-Nordstrom  Black hole charge.
There is no additional $U(1)$ in bulk to modify the geometry
of the AdS black hole. In this paper we follow the more
traditional route of introducing the chemical
potential via the nonperturbative induced DBI plus CS
action through the external $U(1)_V$ source
in Sakai and Sugimoto chiral model. Specifically, we
introduce the baryon chemical potential as ${\cal V}_0=-i\mu_B/N_c$  where
${\cal V}_\mu$ is the external vector field in the induced DBI plus CS action.
While ${\cal V}_0$ drops in the DBI part it contributes to the CS part
through the WZW as it should. A redefinition of the vector fields
in the Sakai-Sugimoto model necessary to enforce vector dominance,
causes a reshuffling of ${\cal V}_0$ from the CS to the DBI action
as we detail below, with important phenomenological consequences.

In section 2 we review the emergence of the induced effective action
on the probe D8 brane from the 5 dimensional D4-brane background, and
we detail the contributions of the DBI action to fourth
order in the vector fields, as well as the Chern-Simons part. The
expanded effective action enforces total vector dominance. In section
3 we introduce the baryon chemical potential and details the nature of
the effective expansion at finite density. In section 4, we extract
the baryon number density, the pressure and the energy density of
holographic dense matter. In section 5, we discuss the pion and
leading vector meson parameters and interactions in holographic
dense matter. Our conclusions and suggestions are summarized in
section 6.

\section{Duality and Branes}

The {\it generalized holographic principle} states that at
large $N_c$ maximally supersymmetric and conformal gauge
theories in flat D-dimensions are dual to Superstring/Mtheory on
pertinent AdS spaces, i.e. for $D=4$ it is $AdS_5\times S^5$
and for $D=6$ it is $AdS_7\times S^4$.

To model QCD in
$D=4$ dimensions, we start from $D=6$ superconformal
theory, and compactify it twice. First through $S^1$ with
radius $R_1$ leading to $D=5$ SUSY theory. Second through another
$S^1$ of radius $R_2\gg R_1$ with fermion antiperiodic boundary conditions
to break SUSY. Below the cutoff scale $M_{KK}=1/R_2$, the D=5
broken SUSY theory behaves as a D=4 gauge theory.

\subsection{D4/D8 Branes}

Doubly compactified D=6 is the boundary of $AdS_7$ in which
the M-theory reduces to type IIA string theory. As a result
M5 branes wrapping around the first $S^1$ with radius $R^1$
transmute to D4-branes. $N_c$ copies of these branes yield
near extremal black-hole background.

The metric, dilaton $\phi$ and the RR three-form field
$C_3$ in a D4-brane background are given by

\begin{eqnarray}
&&ds^2=\left(\frac{U}{R}\right)^{3/2}
\left(\eta_{\mu\nu}dx^\mu dx^\nu+f(U)d\tau^2\right)
+\left(\frac{R}{U}\right)^{3/2}
\left(\frac{dU^2}{f(U)}+U^2 d\Omega_4^2\right),
\nn\\
&&~~~~e^\phi= g_s \left(\frac{U}{R}\right)^{3/4},
~~F_4\equiv dC_3=\frac{2\pi N_c}{V_4}\epsilon_4 \ ,
~~~f(U)\equiv 1-\frac{\Ukk^3}{U^3} \ .
\label{D4sol}
\end{eqnarray}
Here $\mu=0,1,2,3$ and $\tau$ is the compact variable on $S^1$.
$U>U_{KK}$ is the radial coordinate along $56789$, $\epsilon_4$
the 4-form and $V_4=8\pi^2/3$ the volume of a unit $S^4$
surrounding the D-4 brane. $R^3 \equiv \pi g_s N_c l_s^3$, where
$g_s$ and $l_s$ are the string coupling and length respectively.
Again, this background represents
$N_c$ D-4 branes wrapped on $S^1$.

Let us consider $N_f$ $D8$-probe-branes in this background, which may be
described by $ \tau (U) $. We choose a specific configuration such that
$ \tau $ is constant($\tau = \frac{\pi}{4 M_{KK}}$). It corresponds to
the maximal asymptotic separation between $D8$ and $\bar{D} 8$.
with a new variable $z$, instead of $U$, defined by the relation,
\begin{eqnarray}
U = (U^3_{KK} + U_{KK}z^2)^{\frac{1}{3}} \equiv U_z
\end{eqnarray}
the induced metric on D8 is
\begin{eqnarray}
ds^2_{D8} &=& g_{MN}dx^Mdx^N \nn \\
&=&\left(\frac{U_z}{R}\right)^{3/2}
\eta_{\mu\nu}dx^\mu dx^\nu+
\frac{4}{9}\left(\frac{R}{U_z}\right)^{3/2}\frac{U_{KK}}{U_z}dz^2
+\left(\frac{R}{U_z}\right)^{3/2}U_z^2 d\Omega_4^2 \ ,
\end{eqnarray}
where
\begin{eqnarray}
M(N) &=& \{ \mu(\nu)(0,1,2,3),z(4),\alpha(5,6,7,8) \}
\end{eqnarray}

\subsection{DBI and CS Action on D8}

Consider the $U(N_f)$ gauge field $A_M$ on the probe D8-brane
configuration. The effective action is 9-dimensional and
reads

\begin{eqnarray}
S_{D8}&=& S_{DBI} + S_{CS}\ ,\nn \\
S_{DBI}&=&
-T \int d^9 x \tr \left(\,e^{-\phi}
\sqrt{-\det(g_{MN}+2\pi\alpha' F_{MN})}\right) \\
S_{CS}&=&\frac{1}{48\pi^3} \int_{D8} C_3 \tr F^3
\label{CS1} =\frac{1}{48\pi^3} \int_{D8} F_4\, \omega_5(A) \ ,
\label{CS2}
\end{eqnarray}
where $e^{-\phi} = g_s \left(\frac{U_z}{R}\right)^{-\frac{3}{4}}$,
$F_4=d C_3$ is the RR 4-form field strength and $\omega_5(A)$
is the Chern-Simons 5-form,

\begin{eqnarray}
\omega_5(A)=\tr\left(
AF^2-\half A^3 F+\frac{1}{10}A^5
\right) \ ,
\end{eqnarray}
which satisfies $d\omega_5=\tr F^3$.

Assuming that $A_\alpha = 0$ and $A_\mu$ and $A_z$ are
independent of the coordinates on the $S^4$, the induced
DBI action becomes 5-dimensional

\begin{eqnarray}
S_{DBI}&=& -T \int d^9 x \tr \left(\,e^{-\phi}
\sqrt{-\det(g_{MN}+2\pi\alpha' F_{MN})}\right) \\
&=& -\wt T
\int d^4x dz\, U_z^2 \, \nn \\
& & ~~~~~ \tr \, \sqrt {\, 1 +
(2\pi\alpha')^2 \frac{R^3}{2U_z^3}
\eta^{\mu\nu}\eta^{\rho\sigma} F_{\mu \rho}F_{\nu \sigma}
+(2\pi\alpha')^2 \frac{9}{4}\frac{U_z}{\Ukk}
\eta^{\mu\nu} F_{\mu z}F_{\nu z}
+[F^4]+[F^5]} \ ,\nn\\
\label{FF}
\end{eqnarray}
where $\wt T = \frac{N_c}{216\pi^5}\frac{M_{KK}}{\alpha'^3}$.
$[F^4]$ and $[F^5]$ are short for structures of the type

\begin{eqnarray}
&&[F^4] \sim F_{\mu z}^2 F_{\mu \nu}^2 + F_{\mu\nu}^4  \\
&&[F^5] \sim F_{\mu z}^2 F_{\mu \nu}^3
\end{eqnarray}
The Chern-Simons term is

\begin{eqnarray}
S^{D8}_{CS}=\frac{N_c}{24\pi^2}\, \int_{M^4\times\bR} \omega_5(A) \ ,
\label{D8CS}
\end{eqnarray}
with the normalization $ \frac{1}{2\pi}\int_{S^4} F_4=N_c$,
and $M^4\times \bR$ is the five-dimensional plane parameterized by
$x^\mu$ and $z$.

\subsection{Effective Action in 4-dimension}

In this part we review how the 5-dimendional DBI action yields
the 4-dimensional effective action. Essentially, the 5-dimensional
induced action with compact $S^1$ is 4-dimensional for all excitations
with wavelengths larger than the compactification radius of the order
of $1/M_{KK}$. Throughout and for zero baryon density, our discussion parallels the original discussion
by Sakai and Sugimoto~\cite{Sakai}. We quote it for notation and completeness,
and refer to their work for further details.

The leading terms in the $1/\lambda\approx 1/M_{KK}$ expansion of
the DBI action is

\begin{eqnarray}
S_{\rm D8}^{\rm DBI}= \kappa\int d^4 x dZ \, \tr\left[\, \half
K^{-1/3} F_{\mu\nu}^2+ K F_{\mu z}^2 \,\right]\ , \label{DBI}
\end{eqnarray}
where
\begin{eqnarray}
\kappa \equiv \wt T (2\pi \alpha')^2 R^3 =\frac{\lambda
N_c}{108\pi^3}\ ,  \ \ Z\equiv \frac{z}{U_{KK}}  \ , \ \ K \equiv
1+Z^2 \ .
\end{eqnarray}

In order to extract four-dimensional meson fields out of the five
dimensional gauge field, we expand the gauge field as

\begin{eqnarray}
A_\mu(x^\mu,z)&=&
\sum_{n=1}^\infty B_\mu^{(n)}(x^\mu)\psi_n(z)\ ,
\label{exp1}\\
A_z(x^\mu,z)&=&\varphi^{(0)}(x^\mu)\phi_0(z)+
\sum_{n=1}^\infty \varphi^{(n)}(x^\mu)\phi_n(z)\ ,
\label{exp2}
\end{eqnarray}
We choose the functions
$\psi_n(z)$ to be the eigenfunctions satisfying the
self-adjoint differential equation

\begin{equation}
-K^{1/3}\,\del_Z\left( K\,\del_Z\psi_n\right)= \lambda_n\psi_n \,
\quad \label{deqn;psi}
\end{equation}
where $\lambda_n$ is the eigenvalue and $\{\psi_n\}$ is a complete set.
the normalization is fixed by the Laplacian in (\ref{deqn;psi})

\begin{eqnarray}
\kappa \int dZ \,K^{-1/3}\psi_n\psi_m&=&\delta_{nm}\  .
\label{norm;psi}
\end{eqnarray}
This implies

\begin{eqnarray}
\kappa \int dZ \,K\del_Z\psi_n\del_Z\psi_m&=&\lambda_n\delta_{nm}
\label{norm;dpsi}
\end{eqnarray}
The $\phi_n(Z)$ are chosen such that

\begin{eqnarray}
\phi_n(Z)  &=& (m_n U_{KK})^{-1} \del_Z\psi_n(Z)\quad (n\ge 1) \nn \\
\phi_0(Z)  &=& \frac{1}{\sqrt{\pi\kappa}M_{kk}U_{KK}K}
\end{eqnarray}
satisfying the normalization condition:
\begin{eqnarray}
(\phi_m,\phi_n)\equiv
\frac{9}{4}\,\wt T(2\pi\alpha')^2\Ukk^3
\int dZ \,K\,\phi_m \phi_n  =\delta_{mn} \ ,
\end{eqnarray}
which is compatible with (\ref{deqn;psi}) and (\ref{norm;psi}).

Inserting (\ref{exp1}) and (\ref{exp2}) into
(\ref{DBI}) and using the
orthonomality of $\psi_n$ and $\phi_n$, yield

\begin{eqnarray}
S_{DBI}&=&
\int d^4 x\,\tr\left[
(\del_\mu\varphi^{(0)})^2+\sum_{n=1}^\infty\left(
\half
(\del_\mu B^{(n)}_\nu-\del_\nu B^{(n)}_\mu)^2
+\lambda_nM_{KK}^2
(B_\mu^{(n)}-\lambda_n^{-1/2}\del_\mu\varphi^{(n)})^2
\right)\right]\nn\\
&&~~~~~+(\mbox{interaction terms})\ .
\end{eqnarray}
In the expansion (\ref{exp1}) and (\ref{exp2}), we have
implicitly assumed that the gauge fields are zero asymptotically,
i.e. $A_M(x^\mu,z)\ra 0$ as $z\ra\pm\infty$. The residual gauge
transformation  that does not break this condition is obtained by
a gauge function $g(x^\mu,z)$ that asymptotes a constant
$g(x^\mu,z)\ra g_\pm$ at $z\rm \pm\infty$.
We interpret $(g_+,g_-)$ as elements of the chiral symmetry
group $U(N_f)_L\times U(N_f)_R$ in QCD with $N_f$ massless flavors.

\subsubsection{External photons}

By weakly gauging the $U(N_f)_L\times U(N_f)_R$ chiral
symmetry, we may introduce the external gauge fields
$(A_{L\mu},A_{R\mu})$. For the interaction between
mesons and photon $A^{\rm em}_\mu$, we may choose

\begin{eqnarray}
A_{L\mu}=A_{R\mu}= e Q A_\mu^{\rm em}\ ,
\end{eqnarray}
where $e$ is the electromagnetic coupling constant and
$Q$ is the electric matrix-valued charge

\begin{eqnarray}
Q=\frac{1}{3}\left(
\begin{array}{ccc}
2\\
&-1\\
&&-1
\end{array}
\right)\ ,
\end{eqnarray}
for $N_f=3$.
To insert the external-source gauge fields, we impose
the asymptotic values of the gauge field $A_\mu$ on the
D8-probe-brane as

\begin{eqnarray}
\lim_{z\ra+\infty}
A_\mu(x^\mu,z)= A_{L\mu}(x^\mu)\ ,~~~
\lim_{z\ra-\infty}
A_\mu(x^\mu,z)= A_{R\mu}(x^\mu)\ .
\label{asymp}
\end{eqnarray}
This is implemented by modifying the mode expansion (\ref{exp1}) as
\begin{eqnarray}
A_\mu(x^\mu,z)&=&
A_{L\mu}(x^\mu)\psi_+(z)+A_{R\mu}(x^\mu)\psi_-(z)+
\sum_{n=1}^\infty B_\mu^{(n)}(x^\mu)\psi_n(z)\ ,
\label{extexp}
\end{eqnarray}
where $\psi_\pm(z)$ are defined as
\begin{eqnarray}
\psi_\pm(z)\equiv\half(1\pm\psi_0(z))\ ,~~~
\psi_0(z)\equiv \frac{2}{\pi}\arctan z \ ,
\end{eqnarray}
which
are the non-normalizable zero modes of (\ref{deqn;psi})
satisfying $\del_z\psi_\pm(z)\propto\phi_0(z)$.

\subsubsection{$A_z = 0$ gauge}

The $A_z=0$ gauge can be achieved by applying the gauge transformation
$A_M\ra g A_M g^{-1}+ g\del_M g^{-1}$
with the gauge function
\begin{eqnarray}
g^{-1}(x^\mu,z)= P \exp\left\{-
\int_{0}^z dz'\, A_z(x^\mu,z')
\right\} \ .
\label{ginv}
\end{eqnarray}
The asymptotic values of (\ref{asymp}) change to
\begin{eqnarray}
\lim_{z\ra+\infty}
A_\mu(x^\mu,z)= A_{L\mu}^{\xi_+}(x^\mu)\ ,~~~
\lim_{z\ra-\infty}
A_\mu(x^\mu,z)= A_{R\mu}^{\xi_-}(x^\mu)\ ,
\end{eqnarray}
and
\begin{eqnarray}
A_{L\mu}^{\xi_+}(x^\mu)&\equiv&
\xi_+(x^\mu)A_{L\mu}(x^\mu)\xi_+^{-1}(x^\mu)+
\xi_+(x^\mu)\del_\mu\xi_+^{-1}(x^\mu)\ ,\\
A_{R\mu}^{\xi_-}(x^\mu)&\equiv&
\xi_-(x^\mu)A_{R\mu}(x^\mu)\xi_-^{-1}(x^\mu)+
\xi_-(x^\mu)\del_\mu\xi_-^{-1}(x^\mu)\ .
\end{eqnarray}
where $\xi_\pm(x^\mu)\equiv \lim_{z\ra\pm\infty}g(x^\mu,z)$.
The gauge field in the $A_z=0$ gauge can be expanded as

\begin{eqnarray}
A_\mu(x^\mu,z)&=&
A_{L\mu}^{\xi_+}(x^\mu)\psi_+(z)
+A_{R\mu}^{\xi_-}(x^\mu)\psi_-(z)
+\sum_{n=1}^\infty B_\mu^{(n)}(x^\mu)\psi_n(z)\ .
\label{5dpot}
\end{eqnarray}

The residual gauge symmetry in the $A_z=0$ gauge
is given by the $z$-independent gauge transformation.
The residual gauge symmetry $h(x^\mu)\in U(N_f)$ and
the weakly gauged chiral symmetry
$(g_+(x^\mu),g_-(x^\mu))\in U(N_f)_L\times U(N_f)_R$
act on these fields as
\begin{eqnarray}
A_{L\mu}&\ra& g_+A_{L\mu}g_+^{-1}+g_+\del_\mu g_+^{-1}\ ,
\label{tr1}\\
A_{R\mu}&\ra& g_-A_{R\mu}g_-^{-1}+g_-\del_\mu g_-^{-1}\ ,\\
\xi_\pm&\ra& h\,\xi_\pm\, g_\pm^{-1}\ ,\label{xih}\\
B_\mu^{(n)}&\ra& h\,B_\mu^{(n)}\, h^{-1}\ .
\label{tr2}
\end{eqnarray}
Here $\xi_\pm(x^\mu)$ are interpreted as
the $U(N_f)$ valued fields $\xi_{L,R}(x^\mu)$
which carry the pion degrees of freedom in the
hidden local symmetry approach.
Actually the transformation property (\ref{xih}) is
the same as that for $\xi_{L,R}(x^\mu)$
if we interpret $h(x^\mu)\in U(N_f)$ as the hidden local symmetry.
They are related to the $U(N_f)$ valued
pion field $U(x^\mu)$ in the chiral Lagrangian by
\begin{eqnarray}
\xi_+^{-1}(x^\mu)\xi_-(x^\mu)=U(x^\mu)\equiv
e^{2i\Pi(x^\mu)/f_\pi} \ .
\end{eqnarray}
The pion field $\Pi(x^\mu)$ is identical to $\varphi^{(0)}(x^\mu)$
in (\ref{exp2}) in leading order.
\footnote{
Here the pion field $\Pi(x^\mu)$ is a Hermitian matrix,
while $\varphi^{(0)}(x^\mu)$ and
the vector meson fields $B_\mu^{(n)}(x^\mu)$ are anti-Hermitian.
}

If we choose the gauge

\begin{eqnarray}
\xi_-(x^\mu)=1, \quad \xi_+^{-1}(x^\mu)=U(x^\mu)=e^{2i\Pi(x^\mu)/f_\pi}
\end{eqnarray}
then the gauge field is
\begin{eqnarray}
A_\mu(x^\mu,z) &=& U^{-1}(x^\mu)A_{L\mu}(x^\mu)U(x^\mu)\psi_+(z)
+A_{R\mu}(x^\mu)\psi_-(z) \nn \\
&+&  U^{-1}(x^\mu)\del_\mu U(x^\mu)\psi_+(z)
+\sum_{n\ge 1} B_\mu^{(n)}(x^\mu) \psi_n(z)  \label{nonAexp}
\label{ab0}
\end{eqnarray}
An alternative gauge is also

\begin{eqnarray}
\xi_+^{-1}(x^\mu)=\xi_-(x^\mu)=e^{i\Pi(x^\mu)/f_\pi}\ .
\end{eqnarray}
In this gauge, the gauge potential in (\ref{5dpot}) can be
expanded up to quadratic order in fields as
\begin{eqnarray}
A_\mu &=&
\left(\cV_\mu+\frac{1}{2f_\pi^2}[\,\Pi,\del_\mu\Pi\,]
-\frac{i}{f_\pi}[\,\Pi,\cA_\mu\,]\right)
+\left(\cA_\mu+\frac{i}{f_\pi}\del_\mu\Pi
-\frac{i}{f_\pi}[\,\Pi,\cV_\mu\,]\right)\psi_0
+\nn\\
&&~~~+\sum_{n=1}^\infty v^n_\mu\,\psi_{2n-1}
+\sum_{n=1}^\infty a^n_\mu\,\psi_{2n}+\cdots\ ,
\label{exp}
\end{eqnarray}
with
\begin{eqnarray}
\cV_\mu\equiv\half(A_{L\mu}+A_{R\mu})\ ,~~~
\cA_\mu\equiv\half(A_{L\mu}-A_{R\mu})\ ,~~~
v^n_\mu\equiv B_\mu^{(2n-1)}\ ,~~~
a^n_\mu\equiv B_\mu^{(2n)}\ .
\label{VAva}
\end{eqnarray}

\subsubsection{Skyrme action and WZW term}

It is interesting to note with Sakai and Sugimoto~\cite{Sakai} that the induced
effective action expanded to fourth order in the gradients yields
exactly a Skyrme-like action

\begin{eqnarray}
S_{DBI}\Big|_{v^n_\mu=a^n_\mu=\cV_\mu=\cA_\mu=0}
=\int d^4 x\left(\frac{f_\pi^2}{4}\tr\left(U^{-1}\del_\mu U\right)^2+
\frac{1}{32e_S^2}\tr\left[U^{-1}\del_\mu U,U^{-1}\del_\nu U\right]^2
\right) \ ,
\label{Skyrme}
\end{eqnarray}
where the pion decay constant $f_\pi$ and
the dimensionless parameter $e_S$ are fixed

\begin{eqnarray}
f_\pi^2&\equiv&\frac{4}{\pi}\kappa=
\frac{1}{27\pi^4}\lambda N_c\ ,
\label{fpi}
\\
e_S^{-2}&\equiv&\kappa \int\! dz\, K^{-1/3}(1-\psi_0^2)^2
\ .
\label{es}
\end{eqnarray}

Also in the $A_z=0$ gauge,
the Chern-Simons term is generic

\begin{eqnarray}
S_{\rm D8}^{\rm CS}&=&
-\frac{N_c}{24\pi^2}\int_{M^4}\left(
\alpha_4(d\xi_+^{-1}\xi_+,A_L)
-\alpha_4(d\xi_-^{-1}\xi_-,A_R)
\right)+\nn\\
&&~~~+\frac{N_c}{24\pi^2}
\int_{M^4\times\bR}\left(\omega_5(A)
-\frac{1}{10}\tr(gdg^{-1})^5\right) \ ,
\label{WZW}
\end{eqnarray}
where $g$ is the gauge function given in (\ref{ginv})
and the 4-form $\alpha_4$ is

\begin{eqnarray}
\alpha_4(V,A)\equiv -\half\tr\left(
V(AdA+dAA+A^3)-\half VAVA-V^3 A
\right) \ .
\end{eqnarray}
As expected (\ref{WZW})
yields the WZW term in QCD with gauged pion fields.
Indeed, if we omit the vector meson fields $B_\mu^{(n)}$,
we have

\begin{eqnarray}
S_{CS}\Big|_{v^n_\mu=a^n_\mu=0}
=-\frac{N_c}{48\pi^2}\int_{M^4}Z
-\frac{N_c}{240\pi^2}\int_{M^4\times \bR}\tr(gdg^{-1})^5 \ ,
\label{WZW0}
\end{eqnarray}
where
\begin{eqnarray}
Z&=&
\tr[(A_R dA_R+dA_R A_R+A_R^3)(U^{-1}A_LU+U^{-1}dU)-{\rm{p.c.}}]+\nn\\
&&+\tr[ dA_RdU^{-1}A_L U-{\rm{p.c.}}]
+\tr[A_R(dU^{-1}U)^3-{\rm{p.c.}}]+\nn\\
&&+\half\tr[(A_RdU^{-1}U)^2-{\rm{p.c.}}]
+\tr[UA_R U^{-1}A_L dUdU^{-1}-{\rm{p.c.}}]-\nn\\
&&-\tr[A_R dU^{-1}UA_R U^{-1}A_LU-{\rm{p.c.}}]
+\half\tr[(A_RU^{-1}A_LU)^2]\ .
\label{Z}
\end{eqnarray}
Here ``p.c.'' represents the terms obtained by
exchanging $A_L\lra A_R,~U\lra U^{-1}$.

\subsubsection{Vector meson dominance}

The induced DBI action (\ref{DBI}) carries exact vector dominance.
Indeed, by diagonalizing the kinetic terms of the vector
meson fields, vector meson dominance emerges from the underlying
DBI action naturally. To see that, we recall that the gauge fields are

\begin{eqnarray}\label{Amu}
A_\mu&=&\cV_\mu+\cA_\mu\psi_0
+\sum_{n=1}^\infty v^n_\mu\,\psi_{2n-1}
+\sum_{n=1}^\infty a^n_\mu\,\psi_{2n}\ ,
\label{extexp2}\\
A_z&=&-i\,\Pi\, \phi_0\ ,
\label{AzPi}
\end{eqnarray}
and involve both the external vector sources ${\cal V, A}$ and the
dynamical vector fields $V, a$. To diagonalize the kinetic terms of
the dynamical vector fields require the introduction of the
physical tilde vector fields $\tilde{v},\tilde{a}$ which are

\begin{eqnarray}
\wt v_\mu^n&\equiv& v_\mu^n+a_{\cV v^n}\cV_\mu \ ,
\label{wtv}\\
\wt a_\mu^n&\equiv& a_\mu^n+a_{\cA a^n}\cA_\mu\ \,\,,
\label{wta}
\end{eqnarray}
in terms of which the gauge fields now read

\begin{eqnarray}
A_\mu
&=&\cV_\mu\,\psi_v+\cA_\mu\psi_a
+\sum_{n=1}^\infty\wt v^n_\mu\,\psi_{2n-1}
+\sum_{n=1}^\infty\wt a^n_\mu\,\psi_{2n}\ ,
\label{extexp3}
\end{eqnarray}
with
\begin{eqnarray}
&&\psi_v\equiv1-\sum_{n=1}^\infty a_{\cV v^n}\psi_{2n-1}\ ,~~~
\psi_a\equiv\psi_0-\sum_{n=1}^\infty a_{\cA a^n}\psi_{2n}\ .\nn\\
&&a_{\cV v^n}\equiv \alpha^n \equiv \kappa\int\! dz\, K^{-1/3}\psi_{2n-1}\ ,~~~
a_{\cA a^n}\equiv\kappa\int\! dz\, K^{-1/3}\psi_{2n}\psi_0\ ,
\label{aVvaAa}
\end{eqnarray}

The DBI action (\ref{DBI}) in terms of the physical vector fields
to fourth order is

\begin{eqnarray}
&&\kappa\int\!dz\,\tr\left[\half K^{-1/3} F_{\mu\nu}^2\right]\nn\\
&=&\tr\Bigg[\,
\frac{1}{2e^2}\left(
(F_{\mu\nu}^{A_L})^2+(F_{\mu\nu}^{A_R})^2\right)+\nn\\
&&~~~
+\half(\del_\mu \wt v_\nu^n-\del_\nu \wt v_\mu^n)^2
+\half(\del_\mu \wt a_\nu^n-\del_\nu \wt a_\mu^n)^2+\nn\\
&&~~~
+(\del_\mu \wt v_\nu^n-\del_\nu \wt v_\mu^n)
([\wt v^{p\mu},\wt v^{q\nu}]\, g_{v^nv^pv^q}+
[\wt a^{p\mu},\wt a^{q\nu}]\, g_{v^na^pa^q})+\nn\\
&&~~~
+(\del_\mu \wt a_\nu^n-\del_\nu \wt a_\mu^n)
([\wt v^{p\mu},\wt a^{q\nu}] -
[\wt v^{q\nu},\wt a^{p\mu}] )\,g_{v^pa^na^q}+\nn\\
&&~~~+
\half[\wt v^m_\mu,\wt v^n_\nu][\wt v^{p\mu},\wt v^{q\nu}]
\,g_{v^mv^nv^pv^q}
+\half[\wt a^m_\mu,\wt a^n_\nu][\wt a^{p\mu},\wt a^{q\nu}]
\,g_{a^ma^na^pa^q}
+\nn\\
&&~~~
+\left([\wt v^m_\mu,\wt v^n_\nu][\wt a^{p\mu},\wt a^{q\nu}]+
[\wt v^m_\mu,\wt a^p_\nu][\wt v^{n\mu},\wt a^{q\nu}]-
[\wt v^m_\mu,\wt a^p_\nu][\wt v^{n\nu},\wt a^{q\mu}]
\right)
\,g_{v^mv^na^pa^q}
\Bigg]\ .
\end{eqnarray}
Note that all the couplings between the external
gauge fields $(A_L,A_R)$ and the vector meson fields
$(\wt v^n,\wt a^n)$ vanish
in the first term of the effective action (\ref{DBI}).
The second term in the effective action (\ref{DBI})

\begin{eqnarray}
&&\kappa\int\!dz\,\tr\left[ K F_{z\nu}^2\right]\nn\\
&=&\tr\bigg[\,m_{v^n}^2(\wt v^n_\mu-a_{\cV v^n}\cV_\mu)^2+
m_{a^n}^2(\wt a^n_\mu-a_{\cA a^n}\cA_\mu)^2+
(i\del_\mu\Pi+f_\pi\cA_\mu)^2+\nn\\
&&~~~+2i g_{a^mv^n\pi}\,\wt a^m_\mu[\Pi,\wt v^{n\mu}]
-2g_{v^n\pi\pi}\,\wt v^n_\mu[\Pi,\del^\mu\Pi]-
\nn\\
&&~~~-c_{a^na^m}[\Pi,\wt a^n_\mu][\Pi,\wt a^{n\mu}]-
c_{v^nv^m}[\Pi,\wt v^n_\mu][\Pi,\wt v^{n\mu}]
\,\bigg]\ ,
\label{Lsec}
\end{eqnarray}
The mesons couple to the external gauge fields only through
$\wt v^n\ra\cV$ and $\wt a^n\ra\cA$ transitions in
(\ref{Lsec}).

Finally, inserting  (\ref{Amu}) and (\ref{AzPi}) into the
Chern-Simons action gives

\begin{eqnarray}
S_{\rm D8}^{\rm CS}
&=&
-\frac{N_c}{4\pi^2}\frac{i}{f_\pi}\int_{M^4} \tr\Big[\,
\Pi\, dB^ndB^m\, c_{nm}+\nn\\
&&~~~~~~+\Pi\,(dB^mB^nB^p+B^mB^ndB^p)\,c_{mnp}
+\Pi\,B^mB^nB^pB^q\,c_{mnpq}\Big]+\nn\\
&&~~~+
\frac{N_c}{24\pi^2}
\int_{M^4} \tr\Big[\,B^mB^ndB^p\,d_{mn|p}
-\frac{3}{2}B^mB^nB^pB^q\, d_{mnp|q}\Big]\ ,
\end{eqnarray}
where $B^{2n-1}\equiv\wt v^n$, $B^{2n}\equiv\wt a^n$
It shows complete vector meson dominace in the WZW term.
The terms with two or more pion fields are absent.

\section{Induced action at finite density}

For simplicity we work in Euclidean space, in which case the
DBI action reads

\begin{eqnarray}
S_{DBI}^E &=&  \wt T
\int d^4x^E dz\, U_z^2 \, \nn \\
& & ~~~~~ \tr \, \sqrt {\, 1 +
(2\pi\alpha')^2 \frac{R^3}{2U_z^3} F_{\mu \nu}^E F_{\mu \nu}^E
+ (2\pi\alpha')^2 \frac{9}{4}\frac{U_z}{\Ukk}
F_{\mu z}^E F_{\mu z}^E
+[(F^E)^4]+[(F^E)^5]} \nn \ , \\
\label{FFE}
\end{eqnarray}
following prescription.
\begin{eqnarray}
x^0 &\rightarrow& -i\tau \ , \nn \\
A_0 &\rightarrow& iA_0^E  \ , \nn \\
S^E &=& -iS \ .
\end{eqnarray}
From now on we omit the superscript $E$.
From the arguments to follow, we will see that due to
vector dominance no {\it matter term} is generated through
the Chern-Simons part. In a way this is expected;
the anomaly in the Chern-Simons form is an ultraviolet effect
that after regularization shows up in the infrared. Since matter
dwells mostly in the infrared, the scale decoupling insures the
insensitivity of the Chern-Simons terms therefore of the anomaly.
This result is usually referred to as the non-renormalization
theorem.

The density will be introduced through the boundary and via
the external vector field ${\cal V}$ through

\begin{eqnarray}
\cV_\mu = - i \, \mu \, \delta_{\mu0} \, 1_{N_f \times
N_f}, \qquad\quad \cA_\mu = 0 \ .
\label{BOUNDARY}
\end{eqnarray}
The master gauge field now reads from (\ref{Amu})

\begin{eqnarray}
A_\mu &=& -  i \mu \delta_{\mu 0}\,(1-\sum_{n=1}^\infty
\alpha^n\psi_{2n-1}) +\sum_{n=1}^\infty\wt v^n_\mu\,\psi_{2n-1}
+\sum_{n=1}^\infty a^n_\mu\,\psi_{2n}\ , \label{extexp33}
\end{eqnarray}
For calculational convenience below,
we split $\wt v_0^n$ in two parts: a
constant $U(1)$ part ($-i\wt v_{0C}^n$) and the rest ($\overline{v_{0C}^n}$),

\begin{eqnarray}
\wt v_0^n&=& -i v_0^n-i \alpha^n\mu  = -iv_{0C}^n+
\overline{v_{0C}^n} - i\alpha^n\mu \nn \\
&=& \underbrace{-i v_{0C}^n -
i\alpha^n\mu}_{=: - i \wt v_{0C}^n } \ + \overline{v_{0C}^n}, \nn \\
\wt v_i^n&\equiv& v_i^n \nn
\label{wta}
\end{eqnarray}
where $-i v_{0C}^n$ and $\overline{v_{0C}^n}$ are
the constant $U(1)$ part and the rest part of $v_0^n$ respectively.
Thus

\begin{eqnarray}
A_0
&=& -i \mu + \sum_{n=1}^\infty i (\mu \alpha^n-\wt v^n_{0C})\psi_{2n-1}
 +\sum_{n=1}^\infty \overline{v^n_{0C}}\,\psi_{2n-1}
+\sum_{n=1}^\infty a^n_0\,\psi_{2n}\ , \nn \\
A_i &=& \sum_{n=1}^\infty v^n_i\,\psi_{2n-1}
+\sum_{n=1}^\infty a^n_i\,\psi_{2n}\ ,
\label{extexp3}
\end{eqnarray}
Defining,

\begin{eqnarray}
X^{2n-1} &:=&  \mu \alpha^n - \wt v^n_{0C}\nn \\
X^{2n} &:=& 0  \nn \\
\bar{B}^{2n-1}_0 &:=& \overline{v^n_{0C}}
\quad \bar{B}^{2n-1}_i := v^n_i \nn \\
\bar{B}^{2n}_\mu &:=& a^n_\mu
\end{eqnarray}
leads

\begin{eqnarray}
A_\mu(x^\mu,z)&=& -i\mu \delta_{\mu0} +
\sum_{n=1}^\infty \{i\Xn \delta_{\mu 0} +
\Bnmu(x^\mu)\, \} \psin(z) \ ,  \nn \\
A_z(x^\mu,z) &=& -i\Pi(x) \phi_0(z) \ .
\label{Amu1}
\end{eqnarray}

The corresponding field strengths are

\begin{eqnarray}
F_{\mu\nu}
&\!=\!& \sum_{n=1}^{\infty} ( \dmu \Bnnu - \dnu \Bnmu)
\psin +\sum_{n,m=1}^{\infty}[\Bnmu,\Bmnu] \psin \psim \ , \\
F_{z \mu}&\!=\!& i \partial_\mu\Pi\,\phi_0 +
\sum_{n=1}^{\infty}\{i\Xn \delta_{\mu 0} + \Bnmu (x^\mu)\, \}\dpsin
+ i\sum_{n=1}^{\infty} [\Pi,\Bnmu]\psin \phio \label{Fmuz} \ .
\end{eqnarray}
Notice that $i\Xn \delta_{\mu 0}$ does not contribute to $F_{\mu \nu}$.
The additional terms due to $i\Xn \delta_{\mu 0}$ come from $F_{\mu z}$.
They contribute to the induced action through

\begin{eqnarray}
F_{\mu \nu}F_{\mu \nu}
&\!=\!&
(\dmu \Bnnu-\dnu \Bnmu)(\dmu \Bmnu-\dnu \Bmmu)  \psin\psim  \nn \\
&& + \left\{
[\Bnmu,\Bmnu] \ ,  (\dmu \Blnu-\dnu \Blmu)\right\}\psin\psim\psil
\ , \label{Fmunu2} \\
F_{\mu z}F_{\mu z} &\!=\!&-(\del_\mu\Pi\del_\mu\Pi)
\phi_0^2+ \Bnmu\Bmmu\dpsin\dpsim
- \Xn\Xm \dpsin \dpsim \nn \\
&& + \left(i \{ \del_\mu \Pi\ ,  \Bnmu \} -
2 (\del_0 \Pi)\Xn \right) \phi_0 \dpsin + 2 i  \Xn \Bmo \dpsin \dpsim \nn \\
&& + \left( - 2\Xn [\Pi\ ,\Bmo] + i\{\Bnmu\ ,
[\Pi\ ,\Bmmu] \} \right) \dpsin\psim\phio
+ \left\{\dmu \Pi \ , [\Bnmu \ ,  \Pi]
\right\}\phio^2 \psin \ , \label{FmunuFmuz}\\
\ [F^4] &\!=\!& (2\pi\alpha')^4 \frac{9}{4} \frac{R^6}{U_{KK}U_z^2 }
\left\{i(\del_0 \Pi)\phi_0 + (i\Xn +
\Bno)\dpsin \right\}\frac{1}{2}F_{ij}F_{ij}
\left\{i(\del_0 \Pi)\phi_0 + (i\Xm+ \Bmo)\dpsim \right\} \nn \\
\ [F^5] &\sim& (2\pi\alpha')^5\Xn\Xm\dpsin\dpsim
\frac{1}{2}(F_{12}F_{23}F_{31} - F_{31}F_{12}F_{23})
\end{eqnarray}
For notational simplicity we have omitted $\sum_{n=1}^\infty$.

The induced DBI action at finite density can be separated into field
independent and dependent parts $P_0$ and $P_1$ respectively

\begin{eqnarray}
P_0 &\equiv& 1 - (2\pi\alpha')^2 \frac{9}{4}\frac{U_z}{\Ukk}
\Xn\Xm \dpsin\dpsim \nn \\
P_1 &\equiv& (2\pi\alpha')^2 \frac{R^3}{2U_z^3} F_{\mu \nu}F_{\mu \nu}
+ (2\pi\alpha')^2 \frac{9}{4}\frac{U_z}{\Ukk}
(F_{\mu z} F_{\mu z}+\Xn\Xm \dpsin\dpsim)
+[(F)^4]+[(F)^5]
\end{eqnarray}
so that

\begin{eqnarray}
S_{DBI} &=& \wt T
\int d^4x dz\, U_z^2 \,
\tr \, \sqrt {\, P_0 + P_1} \, \nn \\
&=& \wt T \int d^4x dz\, U_z^2 \,\tr \,
\left[ \sqrt {\, P_0} + \half \frac{P_1}{\sqrt{P_0}}
-\frac{1}{8}\frac{P_1^2}{{\sqrt{P_0}}^3} + \cdots \right] \nn \\
&=& S_1(\Xn) + S_2[\Pi, v_\mu ; \Xn] + \cdots \ , \nn
\end{eqnarray}
with

\begin{eqnarray}
S_1(\Xn) &\equiv& \wt T \int d^4x dz\, U_z^2 \tr \, \sqrt {\, 1 -
(2\pi\alpha')^2 \frac{9}{4}\frac{U_z}{\Ukk}
\Xn\Xm \dpsin\dpsim }  \\
S_2[\Pi, v_\mu ; \Xn] &\equiv& \wt T
\int d^4x dz\, U_z^2 \tr \, \left[\half
\frac{P_1}{\sqrt{P_0}} -\frac{1}{8}
\frac{P_1^2}{\sqrt{P_0}^3} + \cdots \right]\ .
\end{eqnarray}

We note that the dependence on the chemical potential $\mu$ in
$S_1$ is of the the type $\sqrt{1-\# \mu^2}$. There is no term
linear in $\mu$(except through the Chern-Simons term which is odd
under $t \rightarrow -t$). The reason is that the QCD partition
function is even under $\mu \rightarrow -\mu$ since the matter
spectrum is symmetric around zero quark virtuality. We believe
that this behavior maybe derived geometrically from a change in
the underlying metric, but we are unable to show it.

The chemical potential $\mu$ as defined in (\ref{Amu}) with
(\ref{BOUNDARY})(i.e.  $\cV_\mu = - i \, \mu \, \delta_{\mu0} \,
1_{N_f \times N_f},\ \cA_\mu = 0$) yields $F_{0z} = 0$. Thus there
is no contribution to the DBI action. The only contribution stems
from the Chern-Simons action as can be checked explicitly.
However the definition (\ref{Amu}) is at odd with VMD as
explained in detail in \cite{Sakai}. In other words the physical
vector field are $\tilde{v}_\mu^n$ defined in (\ref{wtv}) and not
$v_\mu^n$ in (\ref{Amu}). So (\ref{Amu}) has to be substituted by
(\ref{extexp33}). In this case $F_{0z} \ne 0$ and $\mu$
contributes to the DBI action but not to the Chern-Simons action.

\section{Free energy of dense matter}

On the boundary the induced DBI action describes a dense hadronic system
at finite baryon density whereby the baryons are Skyrmions in large $N_c$.
The free energy $\Omega$ is given as

\begin{eqnarray}
\Omega &\equiv& -T \mathrm{ ln }Z_G \nn \\
&=& -T \mathrm{ ln } \left( N' \int \cD[v_\mu]\cD[\Pi]
e^{-S_{DBI}^E}\right) \nn \\
&=& -kT \mathrm{ ln } \left( N' \int
\cD[v_\mu]\cD[\Pi] e^{-(S_1(\X) + S_2[\Pi,v_\mu;\Xn])}\right) \nn \\
&=&  TS_1 -T \mathrm{ ln } \left(\int
\cD[v_\mu]\cD[\Pi] e^{-S_2[\Pi,v_\mu;\Xn])} \right)
\end{eqnarray}
where $Z_G$ is the grand partition function and $N'$ is an irrelevant
matter independent constant. The Gibbs relation

\begin{eqnarray}
\Omega = U -TS -\mu N = -PV
\end{eqnarray}
implies that $\Omega(T=0) = U-\mu N$ is the free energy
at zero temperature and finite baryon density. Ignoring
the meson zero-point contributions (classical limit), the
leading contribution to the free energy per unit volume
(minus the pressure) is

\begin{eqnarray}
\epsilon(\Xn) &\equiv& \frac{\Omega(T=0)}{V_3} =
\frac{S_1(\Xn)-S_1(0)}{\beta V_3} \nn \\
&=& \wt T \int dz\, U_z^2 \tr \, \left( \sqrt {\, 1 -
(2\pi\alpha')^2 \frac{9}{4}\frac{U_z}{\Ukk}
\Xn\Xm \dpsin\dpsim } - 1 \right) \nn \\
&=& \frac{N_f N_c M_{KK}^4 \lambda^3}{27^3 \pi^5} \int dZ\,
K^{\frac{2}{3}}  \, \left( \sqrt {\, 1 -
 \frac{27^3 \pi^5 \Xn\Xm }{N_c M_{KK}^2
\lambda^3}K^{\frac{1}{3}}\dPsin\dPsim } - 1 \right)\ ,
\end{eqnarray}
where $K \equiv 1 + Z^2$, $\lambda \equiv  g_{YM}^2 N_c$,
and $\Psi_n \equiv \sqrt{\kappa}\psi_n(Z) =
\sqrt{\tilde{T} (2\pi \alpha)^2 R^3} \psi_n(Z)$ .\\
In terms of the meson physical parameters fixed at
zero density,\footnote{From this section the subscript $0$ in the physical quantity,
like $f_{\pi 0}$, means the value
at zero density.}

\begin{eqnarray}
g_{YM}^2 N_c &=& f_{\pi 0}^2 \frac{27 \pi^4}{N_c M_{KK}^2} \nn \\
M_{KK}^2 &=& \frac{m_{k0}^2}{\lambda_k} \quad k=1,2,3,\cdots \nn \ ,
\end{eqnarray}
the free energy density becomes

\begin{eqnarray}
\epsilon =\frac{C_1 N_f  f_{\pi 0}^6}{N_c^2 m_{10}^2} \int dZ\,
K^{\frac{2}{3}} \, \left( \sqrt {\, 1 -
  K^{\frac{1}{3}} \frac{m_{10}^4 N_c^2 }{D_1 f_{\pi 0}^6}
\left(\alpha^n\mu - \vnoc \right)
\left(\alpha^m\mu - \vmoc \right)\dot{\Psi}_{2n-1}
\dot{\Psi}_{2m-1}} - 1 \right) \nn
\label{EE}
\end{eqnarray}
with
\begin{eqnarray}
\alpha^n &=& \sqrt{\kappa}\int dZ K^{-1/3}
\Psi_{2n-1} \sim \sqrt{N_c} \ , \nn \\
\mu &=& \frac{\mu_B}{N_c}  , \nn \\
 C_1 &\equiv& \lambda_1 \pi^7 \ , \quad \quad  D_1 \equiv \lambda_1^2 \pi^7 \ .
\end{eqnarray}
The sum over the vector meson-species $m,n$ is subsumed. (\ref{EE})
receives corrections from the quantum loops with mesonic insertions.
In a way (\ref{EE}) captures the deformation of the underlying
D8-probe-brane in the presence of a fixed External ${\cal V}_0$
to leading order. The square root shows that the response is not
analytic, with a subtle branch point singularity to be physically
interpreted below.

In the limit $N_c \gg 1$ but fixed baryon density $\mu_B=\mu N_c$
(low density limit),

\begin{eqnarray}
&&K^{\frac{1}{3}} \frac{m_{10}^4 N_c^2 }{D_1 f_{\pi 0}^6}
\left(\alpha^n\frac{\mu_B}{N_c} -
\vnoc\right)\left(\alpha^m\frac{\mu_B}{N_c} -
\vmoc\right)\dot{\Psi}_{2n-1}\dot{\Psi}_{2m-1}  \nn \\
&\sim& \frac{1}{N_c}(\sqrt{N_c}\frac{1}{N_c}-\vnoc)^2 \sim \frac{1}{N_c^2}
\end{eqnarray}
The free energy density can be Tayor expanded as

\begin{eqnarray}
\epsilon &\approx& -\frac{ N_f M_{KK}^2}{2}
\left(\alpha^n\frac{\mu_B}{N_c} -
\vnoc\right)\left(\alpha^m\frac{\mu_B}{N_c} -
\vmoc\right) \int dZ K \dot{\Psi}_{2n-1}
\dot{\Psi}_{2m-1} \nn \\
&\approx&  -\frac{ N_f}{2}
\sum_{n=1}^{\infty} m_{(2n-1)0}^2
(\sqrt{N_c}\frac{1}{N_c}-\vnoc)^2 \sim \frac{N_f}{N_c}\nn
\end{eqnarray}
which is quadratic in $\mu_B$ and of order $N_f/N_c$ which
is small.

In the limit $N_c \gg 1$ but fixed quark density $\mu=N_c/ \mu_B$
(high density limit),

\begin{eqnarray}
&&K^{\frac{1}{3}} \frac{m_{10}^4 N_c^2 }{D_1 f_{\pi 0}^6}
\left(\alpha^n\mu-\vnoc \right)
\left(\alpha^m\mu-\vmoc \right)\dot{\Psi}_{2n-1}\dot{\Psi}_{2m-1} \nn \\
&\sim& \frac{1}{N_c}(\sqrt{N_c} - \vnoc)(\sqrt{N_c} - \vmoc)  \sim {(N_c)^0}
\end{eqnarray}
so that

\begin{eqnarray}
\epsilon &=& \frac{C_1 N_f  f_{\pi 0}^6}{N_c^2 m_{10}^2} \int
dZ\, K^{\frac{2}{3}}  \, \left( \sqrt {\, 1 -
  K^{\frac{1}{3}} \frac{m_{10}^4 N_c^2 }
{D_1 f_{\pi 0}^6} \left(\alpha^n\mu-\vnoc
\right)\left(\alpha^m\mu-\vmoc
\right)\dot{\Psi}_{2n-1}\dot{\Psi}_{2m-1} } - 1 \right) \nn \\
&\sim& -\frac{ N_f}{2} \sum_{n=1}^{\infty}
m_{(2n-1)0}^2 (X^n)^2  \sim {N_f\,N_c}  \quad (X^n \ll 1)
\end{eqnarray}
which is of order $N_f N_c$.

\begin{figure}[!htbp]
  \begin{center}
  \includegraphics[width=12cm]{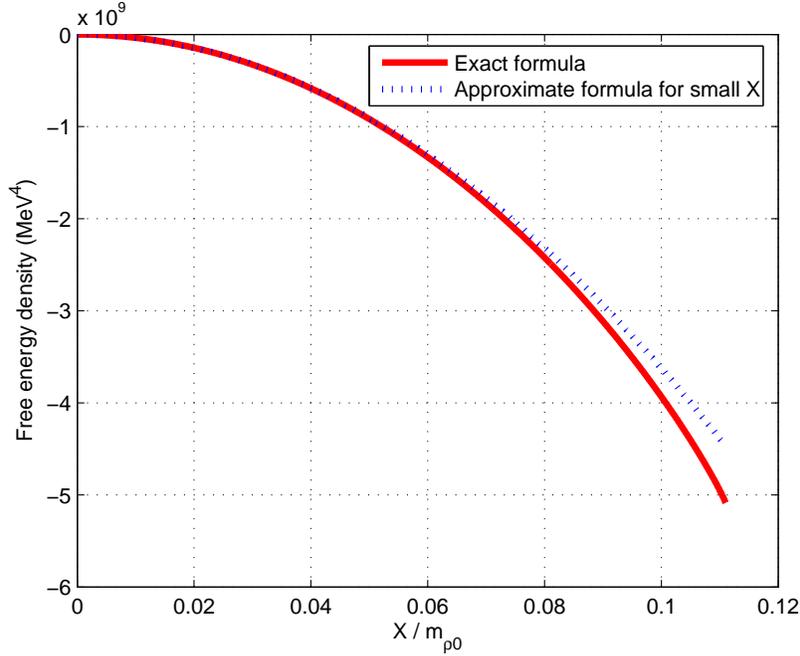}
  \caption {Free energy density vs $\frac{X}{m_{\rho 0}}$}
  \label{grandpotX}
  \end{center}
\end{figure}

In Fig.~\ref{grandpotX} we show the free energy versus
$X^1/m_{\rho 0}$ for $N_c=3, N_f=2$ and $m_{\rho 0} =776\,{\rm
MeV}$, $f_{\pi 0} =93$ MeV. Since \cite{Sakai}
\begin{eqnarray}\label{eigenvalues}
\frac{m_{k0}^2}{M_{KK}^2} = \lambda_k = 0.67,\ 2.88,\ 6.6,\ 11.8,
\ \cdots
\end{eqnarray}
only the $n=m=1$ contribution is justified at low energy. So $X
\equiv X^1$. Here $X$ plays the role of a gauge-shifted chemical
potential.

\begin{figure}[!thbp]
  \begin{center}
  \includegraphics[width=12cm]{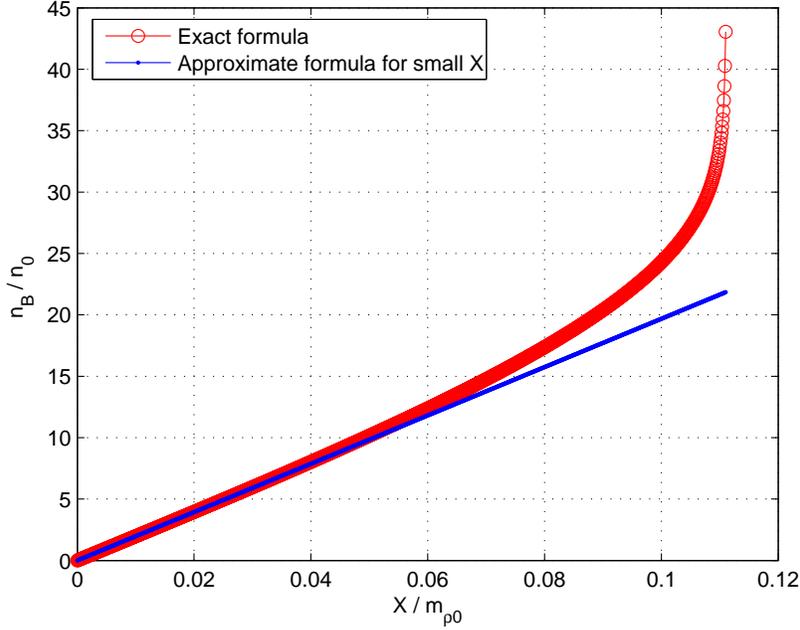}
  \caption {$\frac{n_B}{n_0}$ vs $\frac{X}{m_{\rho 0}}$}
  \label{nBratio}
  \end{center}
\end{figure}

The baryon density $n_B$ follows from the free energy through

\begin{eqnarray}
n_B &:=& \frac{N_B}{V_3} =
-\frac{1}{V_3}\frac{\del \Omega}{\del \mu}  \nn \\
&=& {N_f M_{KK}^2} \int dZ K
\frac{\alpha^1 X^1 \dot{\Psi}_1^2}{
\sqrt{1-K^{\frac{1}{3}} \frac{m_{v0}^4 N_c^2 }
{C_2 f_{\pi 0}^6}(X^1)^2\dot{\Psi}_1^2 }} \nn \\
&\sim& \left\{
       \begin{array}{lll}
          \int dZ \frac{\sqrt{N_c}
\frac{1}{\sqrt{N_c}}}{\sqrt{1-1/N_c^2}}
&\sim N_c^0 & {\mathrm{fixed}} \ \mu_B  \\
          \int dZ \frac{\sqrt{N_c}
\sqrt{N_c}}{\sqrt{1- (N_c)^0}} &\sim N_c & {\mathrm{fixed}}\ \mu
       \end{array} \right.
\label{INTEGRAL}
\end{eqnarray}
where again only the $n=m=1$ contribution was retained. This
is a transcendental equation that reexpresses the shifted chemical
potential $X$ in terms of the physical baryon density $n_B$. For
large $N_c$ and fixed $\mu_B$ it shows that $n_B\sim N_c^0$ which
we interpret as the low density phase. For large $N_c$ and fixed
$\mu=\mu_B/N_c$ it shows that $n_B\sim N_c$ which we interpret
as the high density phase. In Fig.~\ref{nBratio} we show the
numerical solution to (\ref{INTEGRAL}) for $n_B/n_0$ versus
$X/m_{\rho 0}$, where $n_0=0.17\,{\rm fm}^3$ is nuclear matter
density. Asymptotically large densities are attained for a
critical $X/m_{\rho 0} \approx 0.05\,\sqrt{D_1}\,f_{\pi 0} ^3/N_c\,m_{\rho 0}^3$.
For $N_c=3$ and $N_f=2$ we have $X/m_{\rho 0} \sim 0.111$ and $n_B/n_0\sim 43$.

The existence of a limiting chemical potential reflects on the
fact that the underlying D8-probe-brane becomes unstable if the
external ${\cal V}_0$ becomes very large, as the square root in
the ground state energy developes an imaginary part $i\pi$.

\begin{figure}[!htbp]
  \begin{center}
  \includegraphics[width=12cm]{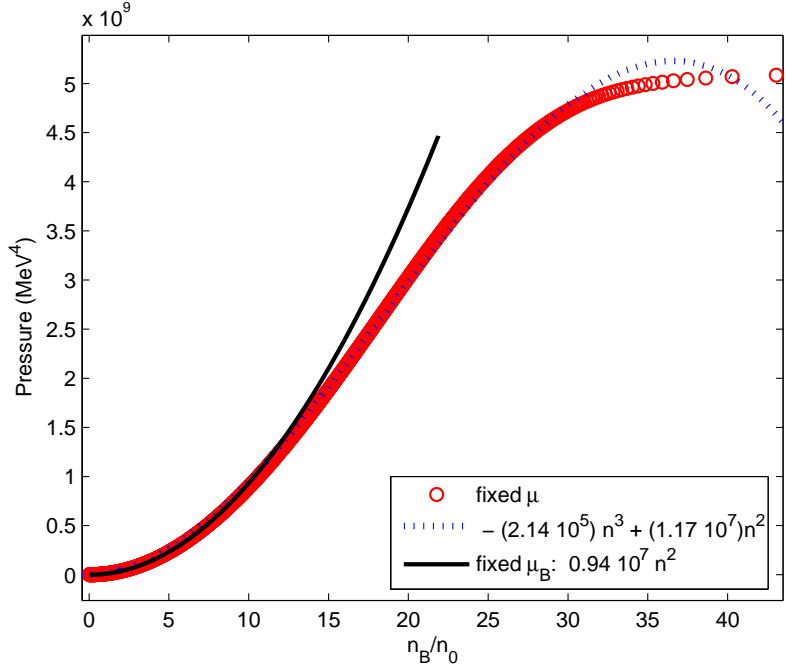}
  \caption {Pressure vs $\frac{n_B}{n_0}$}\label{press}
  \end{center}
\end{figure}

The pressure is $P=-\epsilon$ which is

\begin{eqnarray}
P&=& \frac{C_1 N_f  f_{\pi 0}^6}{N_c^2
m_{\rho 0}^2} \int dZ\, K^{\frac{2}{3}}  \, \left(1- \sqrt {\, 1 -
  K^{\frac{1}{3}} \frac{m_{\rho 0}^4
N_c^2 }{D_1 f_{\pi 0}^6} (\X)^2 \dot{\Psi}_{1}^2 } \right) \nn
\end{eqnarray}
In Fig.~\ref{press} we show how the pressure changes with
baryonic density $n_B/n_0$ for $N_c=3$, $N_f=2$, $m_{\rho 0}=776$
MeV and $f_{\pi 0} =93$ MeV. The pressure is dominated by {\it
repulsive two-body} Skyrmion-Skyrmion like interactions at low
density(the term of order $n^2$ in (\ref{EQ:pressure})). At high
density {\it attractive three-body} Skyrmion-Skyrmion-Skyrmion
like interactions appear and cause the pressure to saturate(the
term of order $n^3$ in (\ref{EQ:pressure})). The two-body
interaction are 100 times stronger throughout the density range
explored. The specific fit is

\begin{eqnarray}\label{EQ:pressure}
P=(1.17\,10^7)\,n^2-(2.14\,10^5)\,n^3
\end{eqnarray}
If we recall that in matter the kinetic energy $K=(\partial P/\partial n)/2$,
then the slope in Fig.~\ref{press} is a measure of the kinetic energy. The
holographic principle suggests that hadronic matter at large $N_c$ freezes
at $n_B\sim 30\,n_0$.

\begin{figure}[!htbp]
  \begin{center}
  \includegraphics[width=12cm]{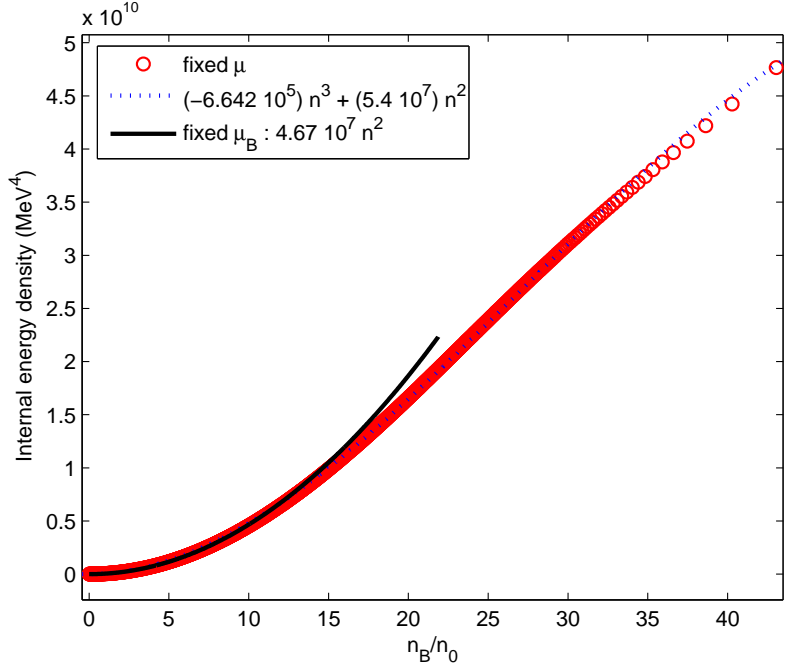}
  \caption { Internal energy density vs $\frac{n_B}{n_0}$}
  \label{eenergy}
  \end{center}
\end{figure}

The energy density is defined as

\begin{eqnarray}
\mathcal{U} &:=& \frac{U}{V_3} = \frac{\Omega + \mu N}{V_3} \nn \\
&=& \epsilon + \mu_B n_B = \epsilon +
\frac{N_c}{\alpha^1}(\voc + X^1) n_B \nn \\
&=& \epsilon + \frac{N_c}{\alpha^1}X^1(n_B)
n_B+\frac{N_c}{\alpha^1} \voc n_B
\end{eqnarray}
and is sensitive to the gauge choice of the {\it dynamical}
$U(1)_v$ vector meson (omega). In the presence of $U(1)_V$
mesons, $\tilde{v}_\mu^n$, the energy density is dependent on the
constant part of the $U(1)_V$ meson field, i.e.
$\tilde{v}_{0C}^n$. This is reminicent of the gauge sensitivity of
the EM potential.
 In the gauge where $\voc=0$, the energy
density is shown in Fig.~\ref{eenergy}. The repulsive three-body
interactions soften the rise in the energy density. Note that the
rise becomes almost linear after $n_B/n_0 \sim 30$. The slope is
sensitive to the choice of the gauge $\voc=0$. Since the pressure
curve shows total freezing at these large densities, we expect
that hadronic matter crystallizes. The slope is expected to be
{\it negative} and a measure of {\it Madelung constant} for the
specific crystal symmetry.

\section{Mesons in Holographic Dense Matter}
The DBI effective action in matter fixes completly the meson
dispersion laws and interactions. Indeed, in matter $S_2[v_\mu ;\Xn]$ is

\begin{eqnarray}
S_2[\Pi, v_\mu ; \Xn] &=& \int d^4x dZ\,
\tr \, \left[ Q \wt T U_{KK} U_z^2 {P_1} -
\frac{1}{4} Q^3 \wt T U_{KK} U_z^2 P_1^2 + \cdots \right]\ .
\end{eqnarray}
where

\begin{eqnarray}
Q(Z;\Xn ) &:=&  \frac{1}{\sqrt{P_0}} =
\frac{1}{\sqrt {\, 1 -
  \Delta K^{\frac{1}{3}}\dot{\Psi}_1^2} }\nn
\end{eqnarray}

\begin{eqnarray}
\Delta := \frac{m_{v0}^4 N_c^2 (\X)^2 }{C_2
f_{\Pi 0}^6} = \frac{27^3 \pi^5 (\X)^2 }{N_c M_{KK}^2 \lambda^3}
\quad \sim \frac{1}{N_c} ~~ \mathrm{fixed} ~~
\mu_B \quad \left(\sim 1~~ \mathrm{fixed}~~ \mu\right)
\end{eqnarray}
Only the term $n=m=1$ in the summation over the vector meson
species was retained (See (\ref{eigenvalues})). In this section we
treat $A_\mu$ as anti-Hermitiaon matrices with generators $t^a$
normalized as $\tr (t^a t^b) = \frac{1}{2}\delta_{ab}$. The
effective action is rotated back to Minkowski space. Throughout,
only terms up to the third order in the fields are retained ,
i.e.$\cO((\Pi,v_{\mu})^4)$. The leading terms to this order in
$[F^5]$ vanishes because of the cyclicity of the trace.

The expansion can be simplified by using the notations

\begin{eqnarray}
\bar{v}_0 := \overline{v^1_{0C}} , \quad X := \X \ , \quad v_i := v_i^1
\end{eqnarray}
so that

\begin{eqnarray}
S_2[\Pi, v_\mu ;\mu]&\!=\!& \int d^4 x \Bigg[ a_{\Pi^2}^T
\tr\left(\del_0\Pi\del_0\Pi\right) -a_{\Pi^2}^S
\tr\left(\del_i\Pi\del_i\Pi\right)\nn \\
&&~~~~ - a_{v^2}^T\tr(\del_0 v_i-\del_i \bvo)^2 +
\frac{1}{2}a_{v^2}^S\tr(\del_i v_j-\del_j v_i)^2 \nn \\
&&~~~~ - {m_v^2}^T \tr \bvo^2 + {m_v^2}^S\tr v_i^2\nn\\
&&~~~~ -a_{v^3}^T \tr\Big( [\bvo,v_i](\del_0 v_i-\del_i \bvo) \Big)
+a_{v^3}^S \tr\Big( [v_i,v_j](\del_i
v_j-\del_j v_i) \Big) \nn \\
&&~~~~ -a_{v\Pi^2}^T \tr\Big(\del_0\Pi[\bvo,\Pi]\Big) +a_{v\Pi^2}^S
\tr\Big(\del_i\Pi[v_i,\Pi]\Big) \nn\\
&&~~~~ - a_{{\bvo}v^2}^T \tr \Big( \bvo (\del_0
v_i-\del_i \bvo)^2 \Big)  + a_{{\bvo}v^2}^S \tr \Big( \bvo (\del_i
v_j-\del_j v_i)^2 \Big) \nn \\
&&~~~~- a_{\bar{v}_0 \Pi^2} (\del_0 \Pi \del_0 \Pi \bvo - \del_0 \Pi \del_i \Pi v_i) - a_{v_0 v^2} \bvo (\bvo^2 - v_i^2)  \nn \\
&&~~~~+\left(\int dZ K Q \phi_0 \dot{\Psi}_1\right) \tr
\left(\frac{2i}{f_\Pi}(\del_\mu \Pi) v_\mu +
\mu \frac{2i}{f_\Pi}(\del_0 \Pi) + \mu [\Pi]^3 \right) \nn \\
&&~~~~ + \left\{\sim \mu (\del_0 \Pi)(\del_i v_j - \del_j v_i)
\int dZ \phi_0 \dot{\Psi}_1 \Psi_1^2 \right\}\Bigg] \ .
\label{pirho1}
\end{eqnarray}
The last two contributions in (\ref{pirho1}) vanish
because $ K$, $Q$, $\Psi_1$ and $
\phi_0$ are even functions and $\dot{\Psi}_1$ is an odd function.
The new two terms related to $\bvo$ are in the 6th and 7th lines of
(\ref{pirho1}).
All the parameters appearing in (\ref{pirho1}) are tabulated below
for non-zero density. Their respective values at zero density are
recorded on the right-most column for comparison. We note that
$Q, \Delta\sim N_c^0$ for fixed $\mu$, so that
all coefficients are of order $N_c^0$ for fixed $\mu$ at large $N_c$.

\begin{center}
\begin{tabular}{|c|l|c|}
\hline
Coefficients & \hspace{2cm}nonzero X & $X=0$($Q=1,\Delta=0)$ \\
\hline
\hline
$a_{\Pi^2}^T$ & $\frac{1}{\pi}\int dZ K^{-1} \, Q
\left(1  +\Delta Q^2 K^{\frac{1}{3}} \dot
\Psi_1^2 \right) $ & $1$ \\
$a_{\Pi^2}^S$ & $\frac{1}{\pi}\int dZK^{-1} \, Q$ & $1$\\
\hline
$a_{v^2}^T$ & $\int dZ\,  K^{-\frac{1}{3}} \Psi_1^2 \, Q $ & $1$\\
$a_{v^2}^S$ & $\int dZ\,  K^{-\frac{1}{3}} \Psi_1^2 \, Q^{-1}  $& $1$\\
\hline
${m_v^2}^T$ & $\frac{m_{v0}^2}{\lambda_1} \int dZ\,
K \dot\Psi_1^2 \, Q\left(1  +\Delta Q^2 K^{\frac{1}{3}} \dot
\Psi_1^2 \right) $ & $m_\rho^2$ \\
${m_v^2}^S$ & $\frac{m_{v0}^2}{\lambda_1}
\int dZ\, K \dot\Psi_1^2 \, Q $ & $m_\rho^2$ \\
\hline
$a_{v^3}^T$  & $\frac{2 m_{v0}}{\sqrt{\pi \lambda_1}f_{\Pi 0}}\,\int
dZ\,K^{-\frac{1}{3}} \Psi_1^3 \,  Q $
& $\frac{2 m_{v0}}{\sqrt{\pi \lambda_1}f_{\Pi 0}} \cdot 0.45 $\\
$a_{v^3}^S$  & $\frac{2 m_{v0}}{\sqrt{\pi \lambda_1}f_{\Pi 0}}\,\int dZ\,
K^{-\frac{1}{3}} \Psi_1^3 \, Q^{-1}  $
& $\frac{2 m_{v0}}{\sqrt{\pi \lambda_1}f_{\Pi 0}} \cdot 0.45 $\\
\hline
$a_{v\Pi^2}^T$ & $\frac{2}{\sqrt{\kappa \pi}}
\int dZ\,   K^{-1} \Psi_1^2 \, Q \left(1  +\Delta Q^2 K^{\frac{1}{3}} \dot
\Psi_1^2 \right)  $ & $\frac{2}{\sqrt{\kappa \pi}}\cdot 0.63 $\\
$a_{v\Pi^2}^S$ & $\frac{2}{\sqrt{\kappa \pi}}
\int dZ\,  K^{-1} \Psi_1^2 \, Q$ & $\frac{2}{\sqrt{\kappa \pi}}\cdot 0.63 $ \\
\hline
\hline
$a_{{v_0}v^2}^T$ & $\frac{\Delta}{2X} \int dZ\,
Q^3  \dot{\Psi}_1^2 \Psi_1^2$ & $0$  \\
$a_{{v_0}v^2}^S$ & $ -{\Delta}{X} \int dZ\,  Q \dot{\Psi}_1^2 \Psi_1^2
 + \frac{\Delta}{2X}
\int dZ\, Q^3  \dot{\Psi}_1^2 \Psi_1^2
+ \frac{\Delta^2}{2X}
\int dZ\, Q^3 K^{\frac{1}{3}} \dot{\Psi}_1^4 \Psi_1^2  \ $ & $0$  \\
$a_{{v_0}\Pi^2}$ &  $ \frac{\Delta}{X}
\int dZ\, Q^3 K^{-\frac{2}{3}} \dot{\Psi}_1^2$ & $0$  \\
$a_{{v_0}v^2}$ &  $ \frac{\Delta}{X}
\int dZ\, Q^3  K^{\frac{4}{3}}\dot{\Psi}_1^4$& $0$ \\
\hline
\end{tabular}
\end{center}

\subsection{Meson Velocities}

\begin{figure}[!hbp]
  \begin{center}
  \includegraphics[width=7cm]{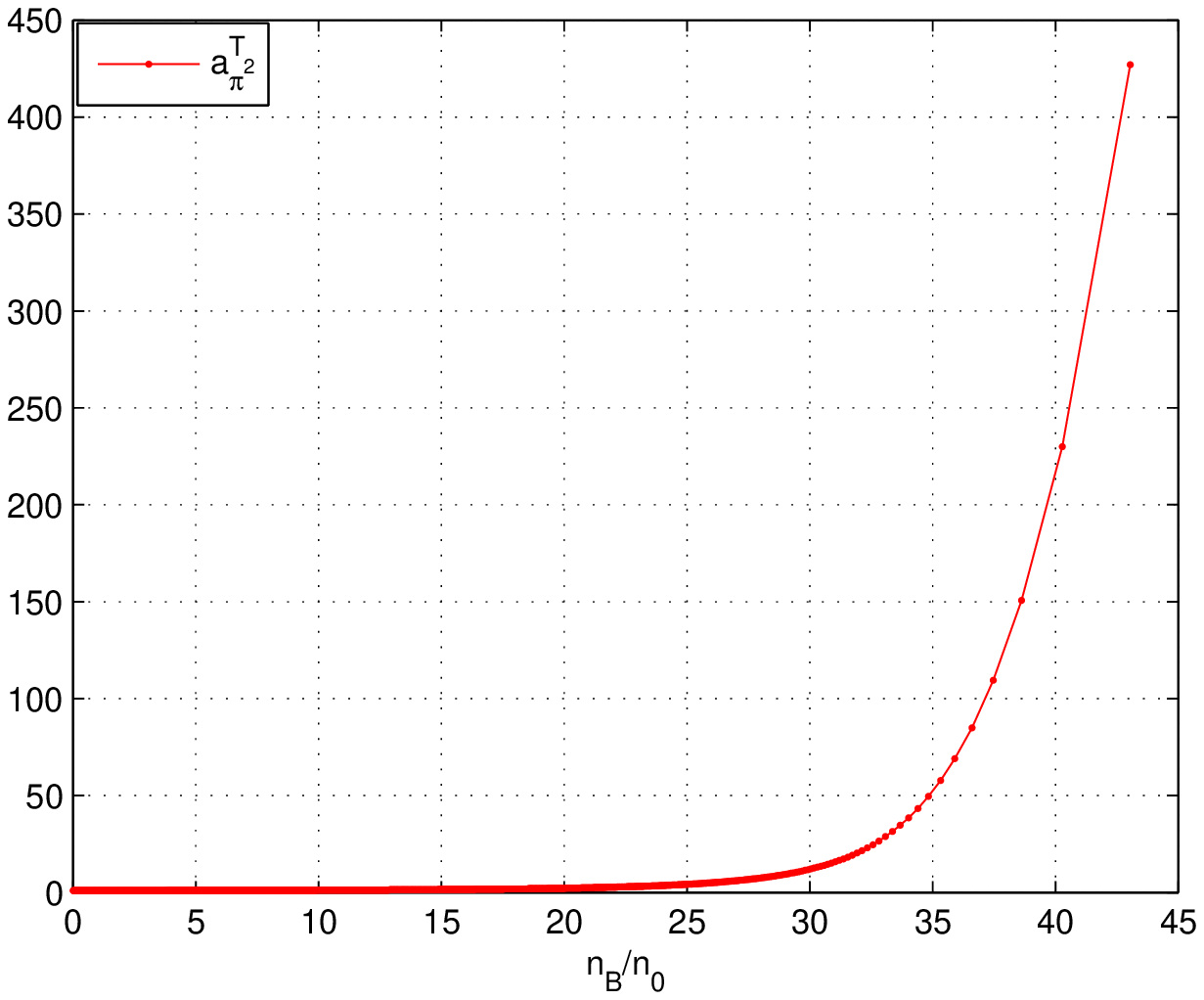}
  \includegraphics[width=7cm]{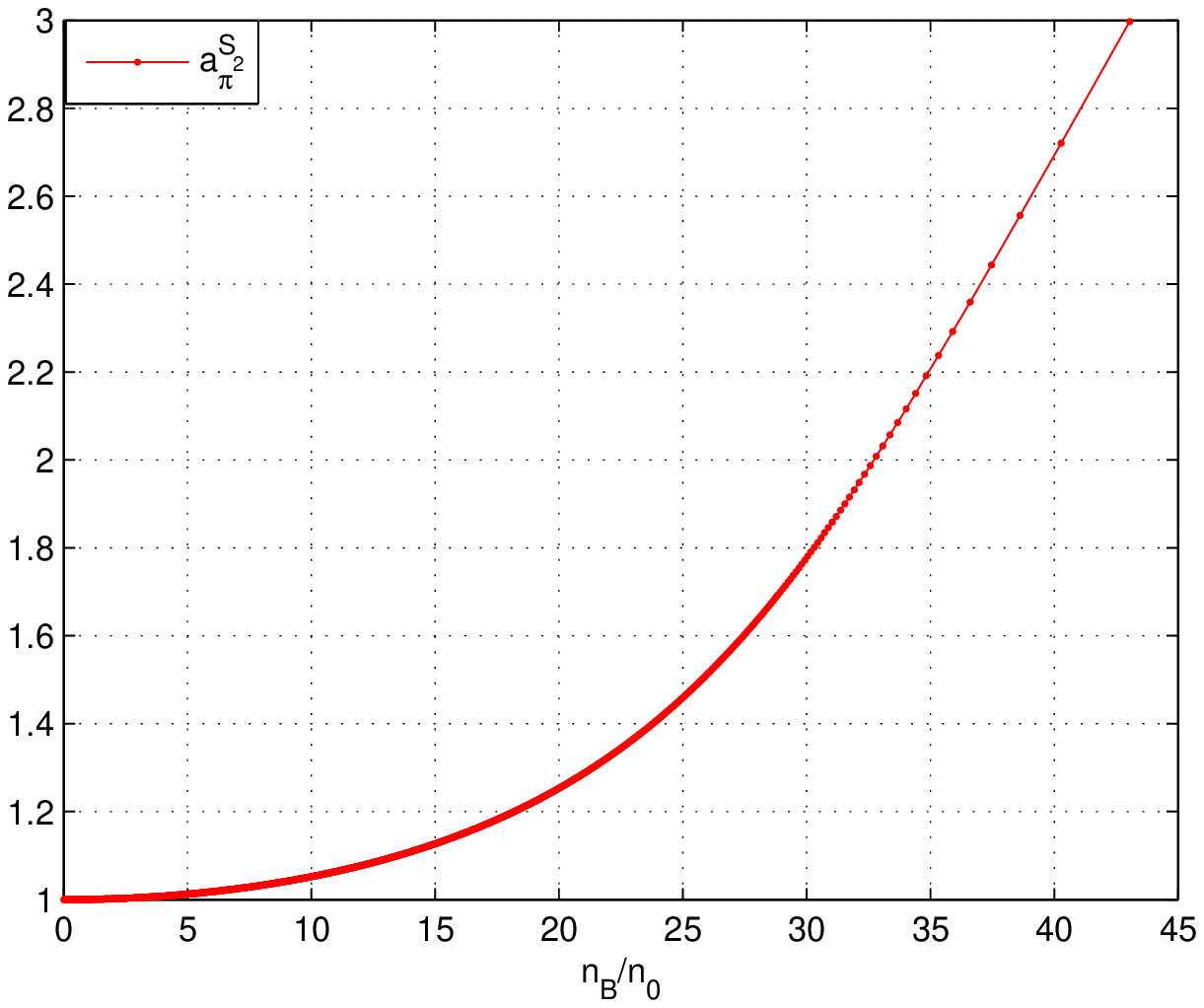}
  \caption{${a_{\pi^2}^T}$ and ${a_{\pi^2}^S}$}
  \label{fpi2tsnormpi}
  \end{center}
\end{figure}

\begin{figure}[!hbp]
  \begin{center}
  \includegraphics[width=7cm]{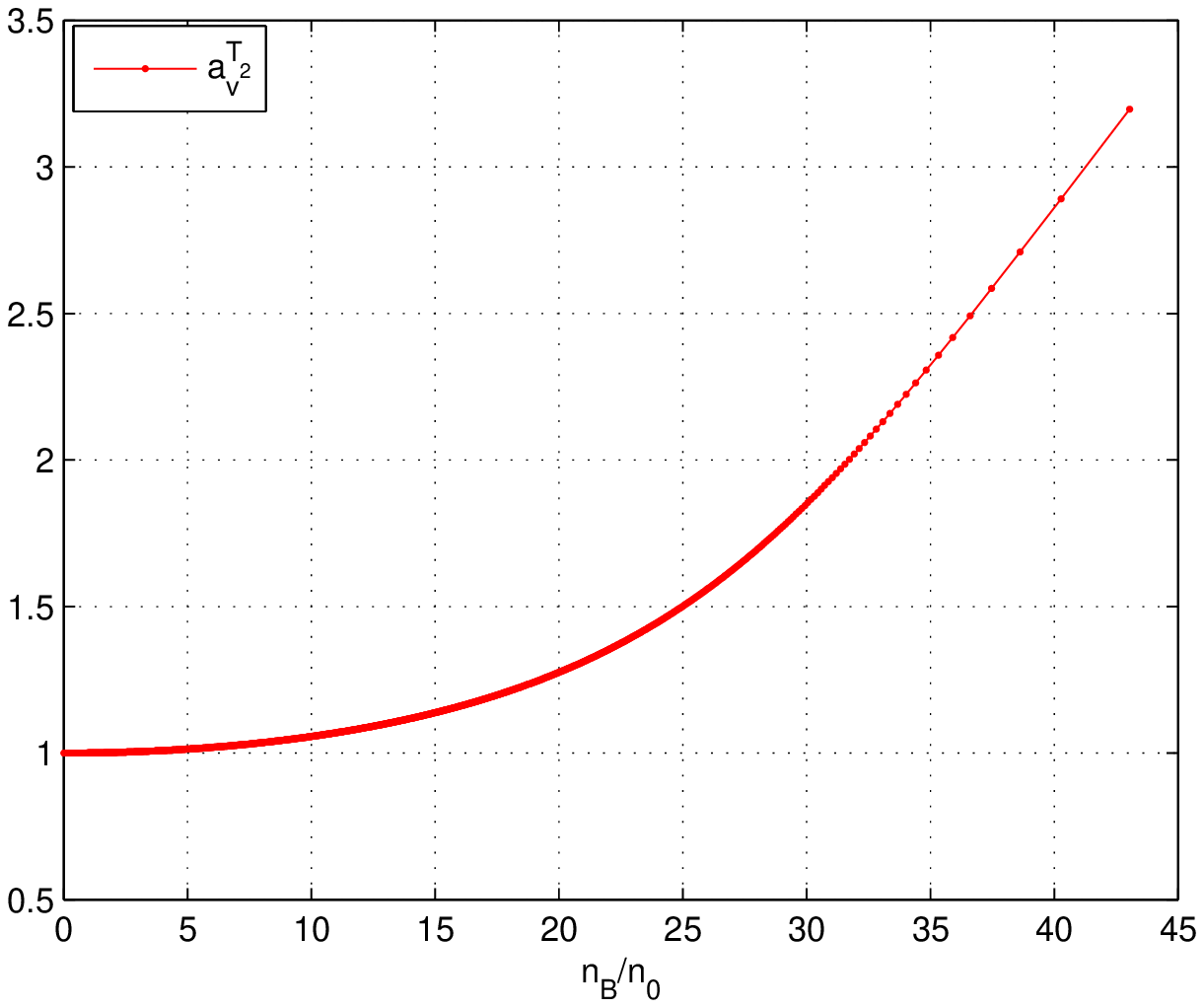}
  \includegraphics[width=7cm]{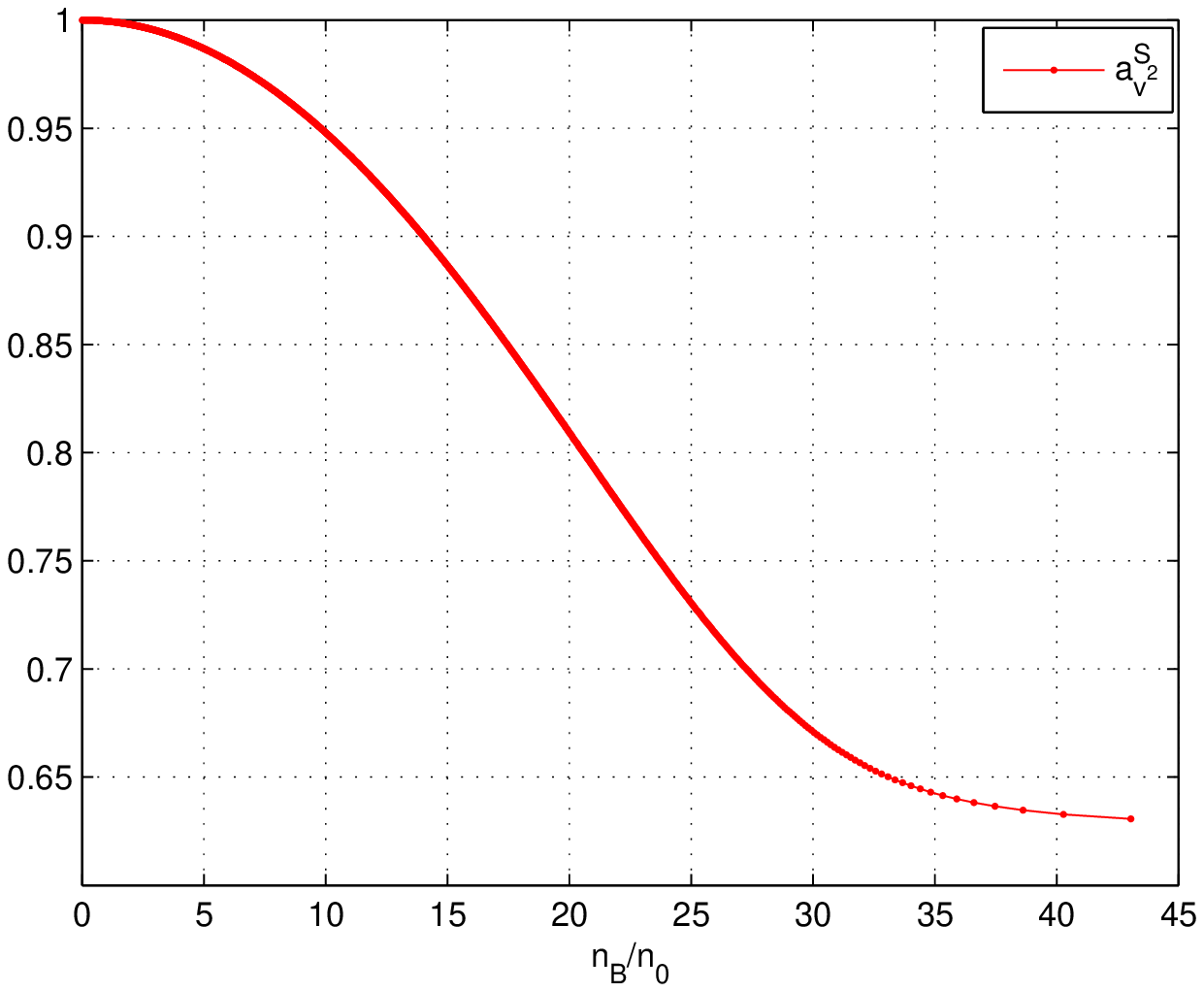}
  \caption{$a_{v^2}^T$ and $a_{v^2}^S$ }
  \label{fpi2tsnormrho}
  \end{center}
\end{figure}

\begin{figure}[!hbp]
  \begin{center}
  \includegraphics[width=7cm]{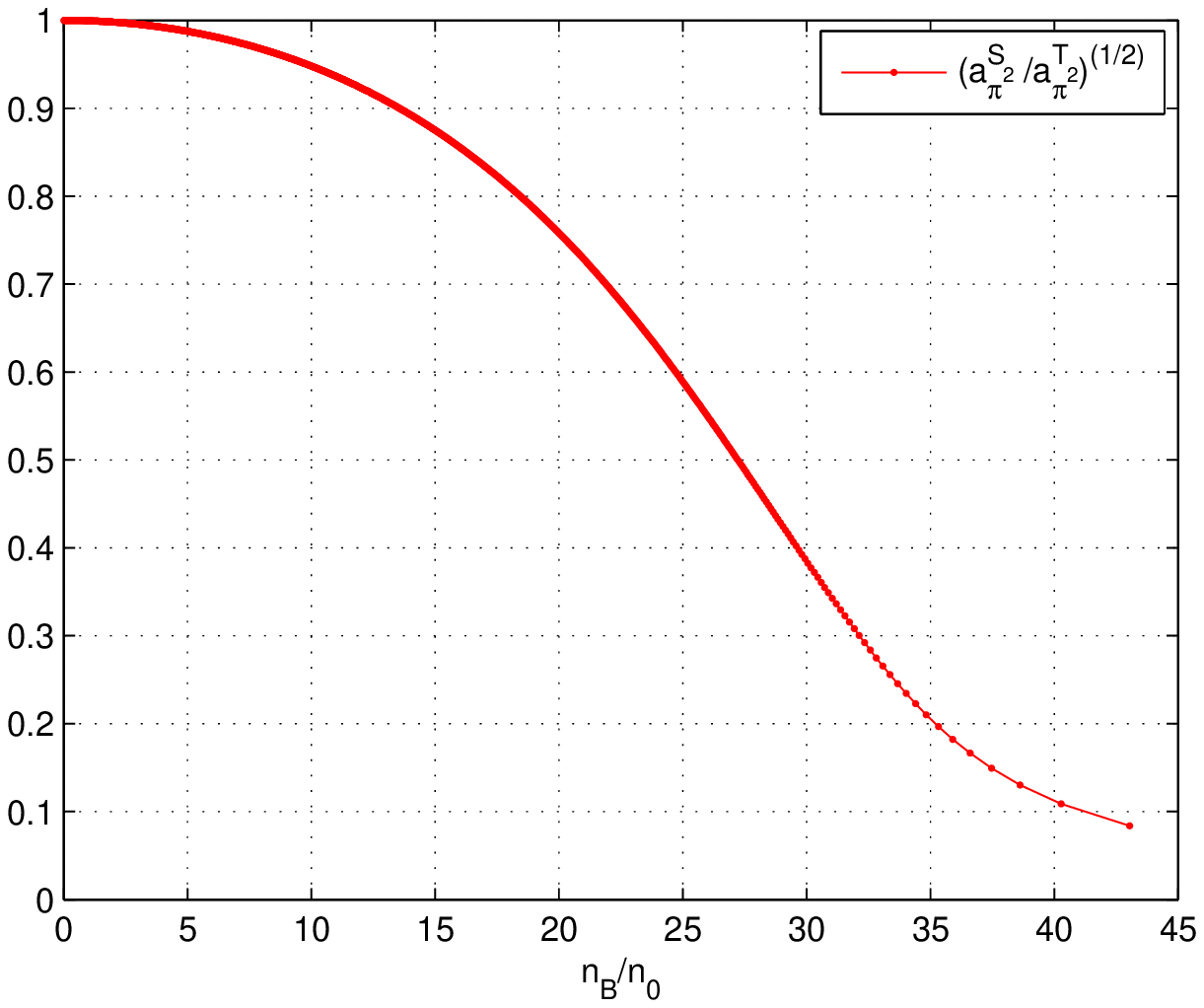}
  \includegraphics[width=7cm]{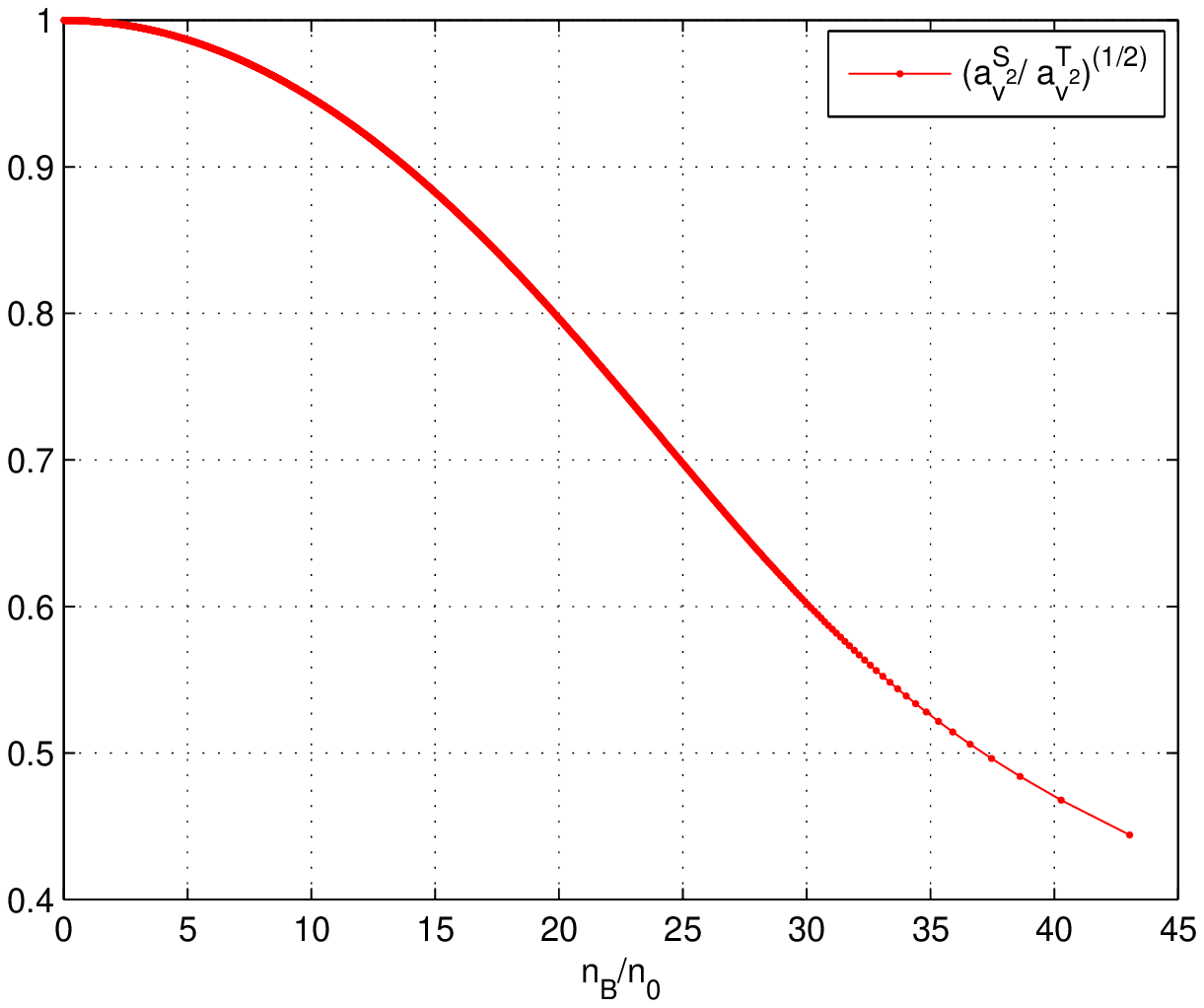}
  \caption{velocity: pion(left) and $\rho$(right)}
  \label{fpi2tsvelo}
  \end{center}
\end{figure}

For $N_c=3$, $m_{vo}=776$MeV, $f_{\pi0}=93$MeV, we show the
pion in Fig.~\ref{fpi2tsnormpi}
and vector meson in Fig.~\ref{fpi2tsnormrho}
renormalization constants, both for
the time and space-components. The resulting velocity of the
pion and the rho meson in holographic dense matter is shown
in Fig.~\ref{fpi2tsvelo}. The pion velocity
\begin{eqnarray} \label{pivelocity}
v_\pi = \sqrt{\frac{a_{\pi^2}^S}{a_{\pi^2}^T}} = \frac{f_\pi^S}{f_\pi^T}
\end{eqnarray}
with $f_\pi^{S,T}$, the spatial ($S$) and temporal ($T$) pion decay constants,
approaches zero when hadronic matter freezes. The vector meson velocity
\begin{eqnarray}
v_v =\sqrt{\frac{a_{v^2}^S}{a_{v^2}^T}}
\end{eqnarray}
is about half.


\subsection{Vector Screening Masses}

\begin{figure}[!hbp]
  \begin{center}
  \includegraphics[width=7cm]{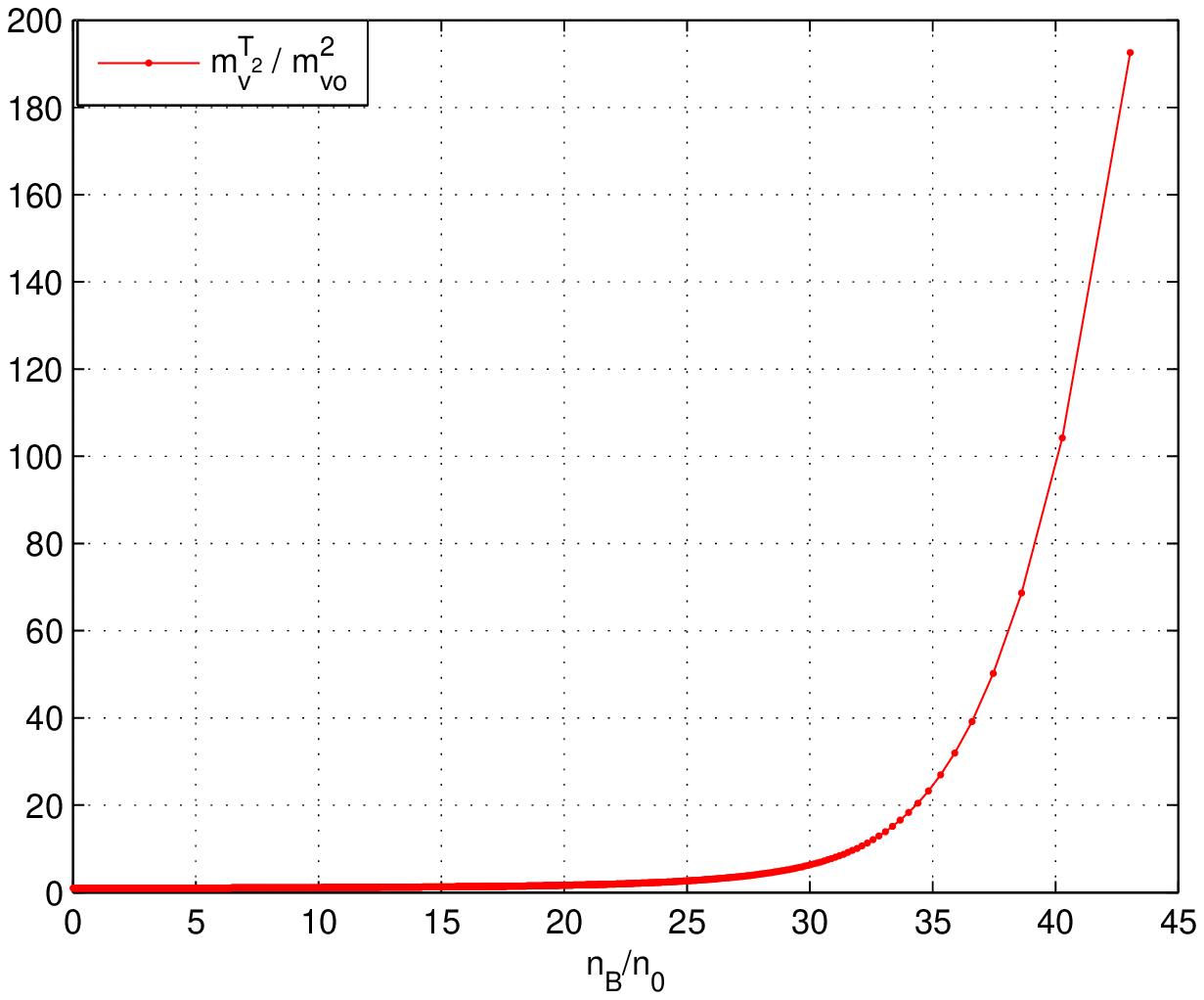}
  \includegraphics[width=7cm]{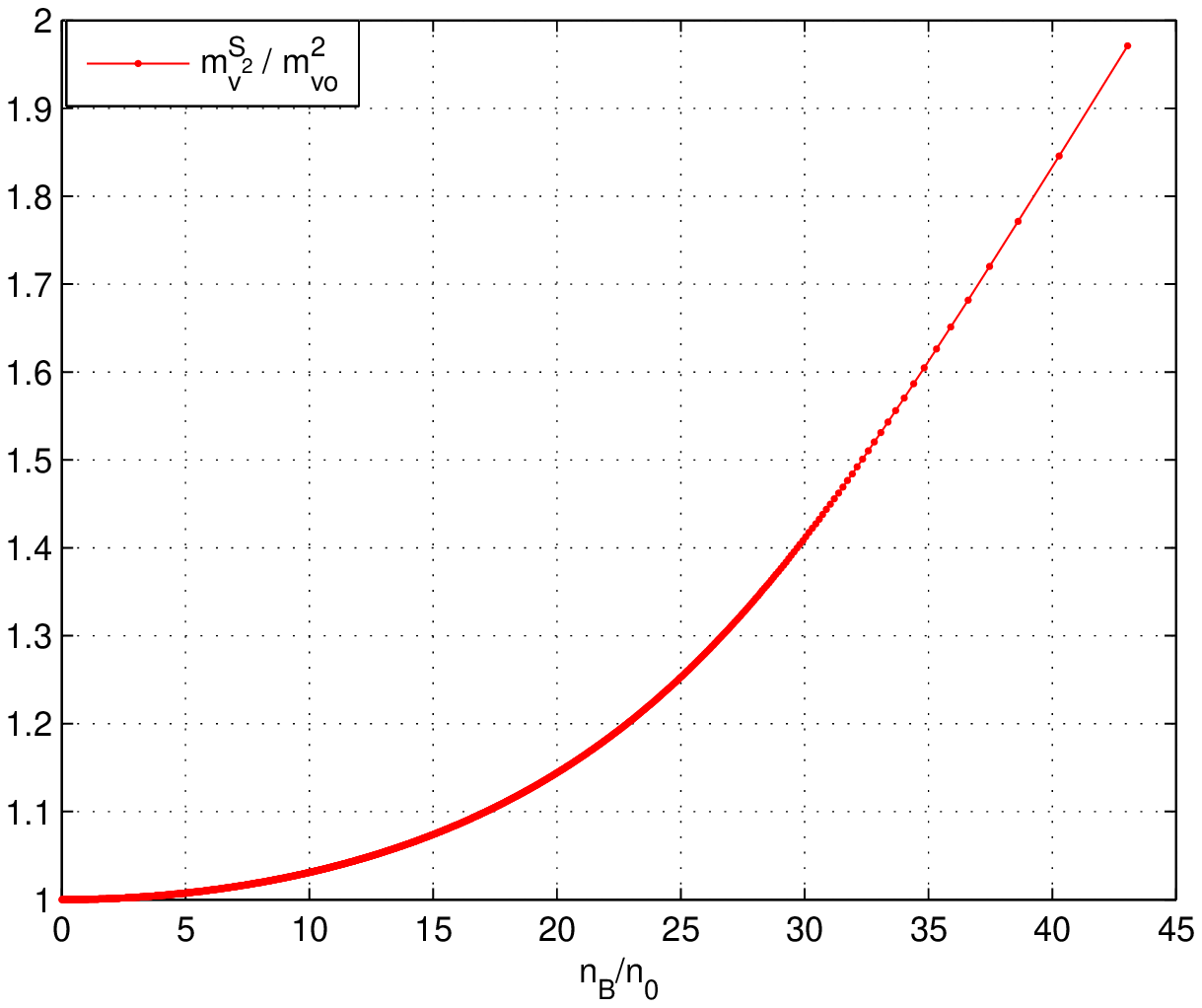}
  \caption{$\frac{{m_v^2}^T}{m_{v0}^2}$ and $\frac{{m_v^2}^S}{m_{v0}^2}$ }
  \label{fpi2tsunr}
  \end{center}
\end{figure}

\begin{figure}[!hbp]
  \begin{center}
  \includegraphics[width=7cm]{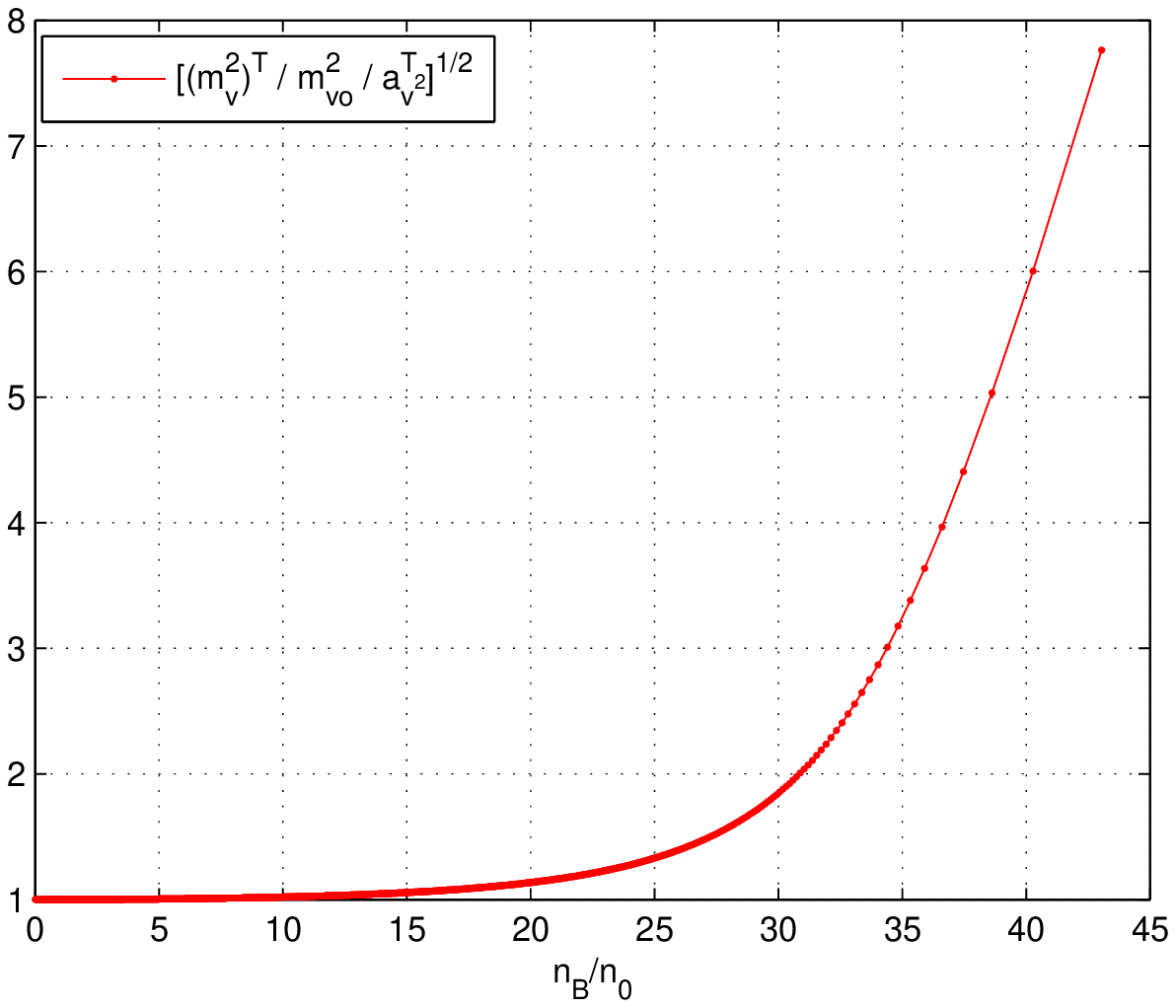}
  \includegraphics[width=7cm]{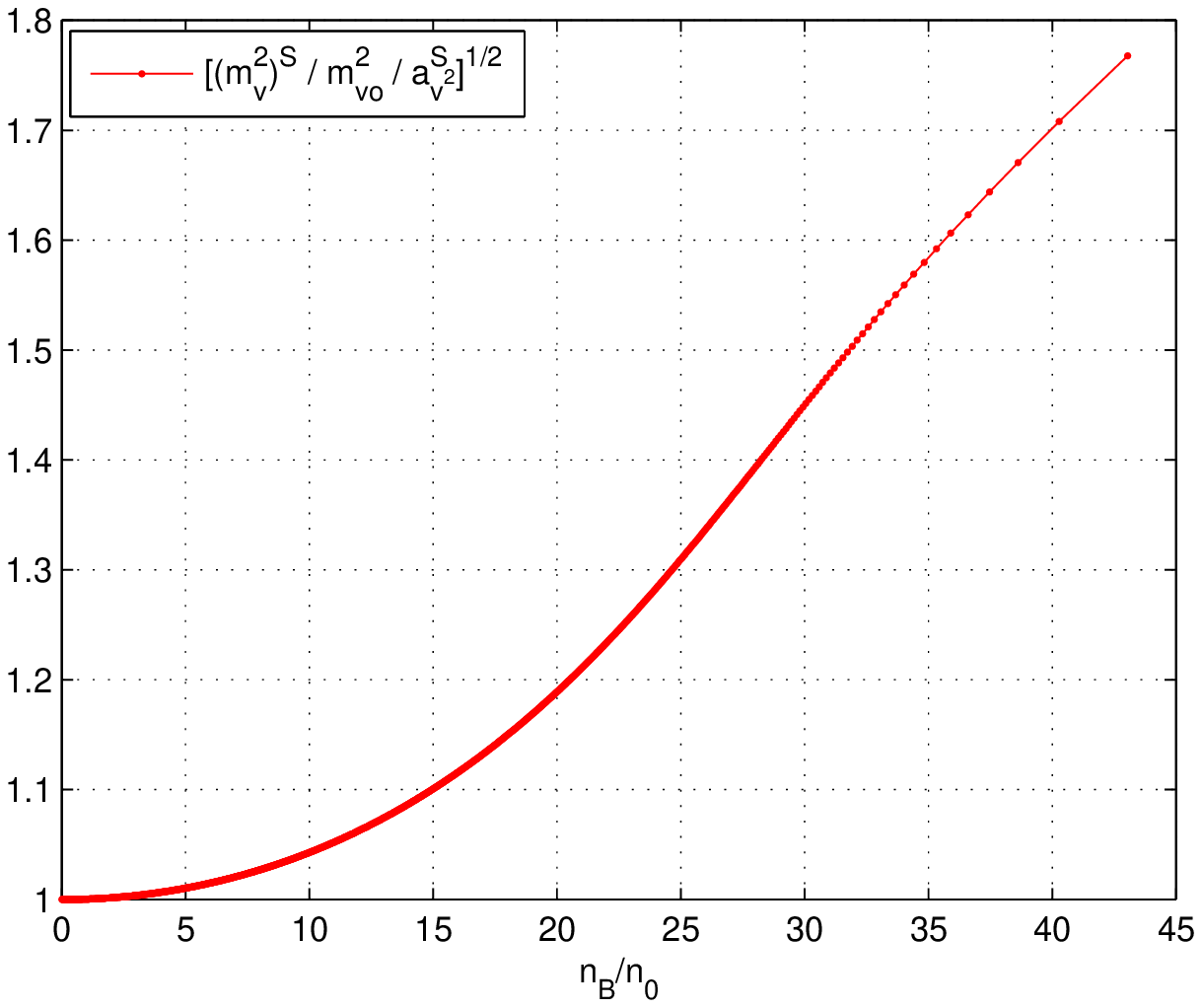}
  \caption{Screening masses: Longitudinal (left) and Transverse (right)}
  \label{fpi2tsscreening}
  \end{center}
\end{figure}

The pion is massless throughout as the current quark masses
are set to zero in this work. The vector meson masses follow
by Higgsing, and change in matter both space-like and time-like.
Space-like, the vector meson masses are screening masses. The
unrenormalized masses increase with increasing baryon density
as shown in Fig.~\ref{fpi2tsunr}. The physical screening masses($M_T$)
are obtained by normalizing with the pertinent matter dependent
{\it space-like} wavefunction renormalizations. Both the vector and
isovector screening masses increase with increasing baryon density
as they should. Fig.~\ref{fpi2tsscreening} show the screening masses
for longitudinal (left) and transverse (right) vectors. They are defined
as

\begin{eqnarray}
M_L^2=\frac{m_v^{2T}/m_{v0}^2}{a_{v^2}^T} \ ,\quad \quad
M_T^2=\frac{m_v^{2S}/m_{v0}^2}{a_{v^2}^S} \ .
\end{eqnarray}


\subsection{Vector Meson Masses}

\begin{figure}[!hbp]
  \begin{center}
  \includegraphics[width=7cm]{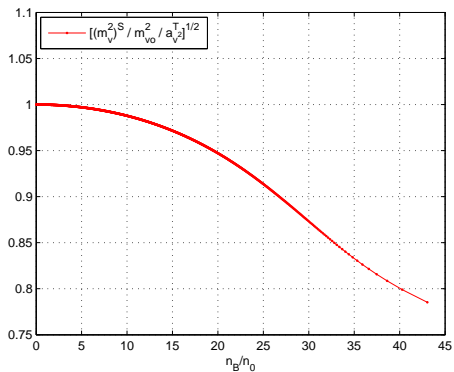}
  \caption{Time-like Vector mass vs baryon density}
  \label{fpi2tsmass}
  \end{center}
\end{figure}

The time-like vector masses are obtained similarly to the screening
masses by instead using the {\it time-like} wavefunction renormalizations.
That is

\begin{eqnarray}
M_V^2=\frac{m_v^{2S}/m_{v0}^2}{a_{v^2}^T}
\end{eqnarray}
The dependence of $M_V^2$ on the baryon density is shown in
Fig.~\ref{fpi2tsmass}. The rho meson mass drops by 20\% when
holographic dense matter freezes.

\subsection{$V\pi\pi$ and $VVV$ Couplings in Matter}

\begin{figure}[!hbp]
  \begin{center}
   \includegraphics[width=7cm]{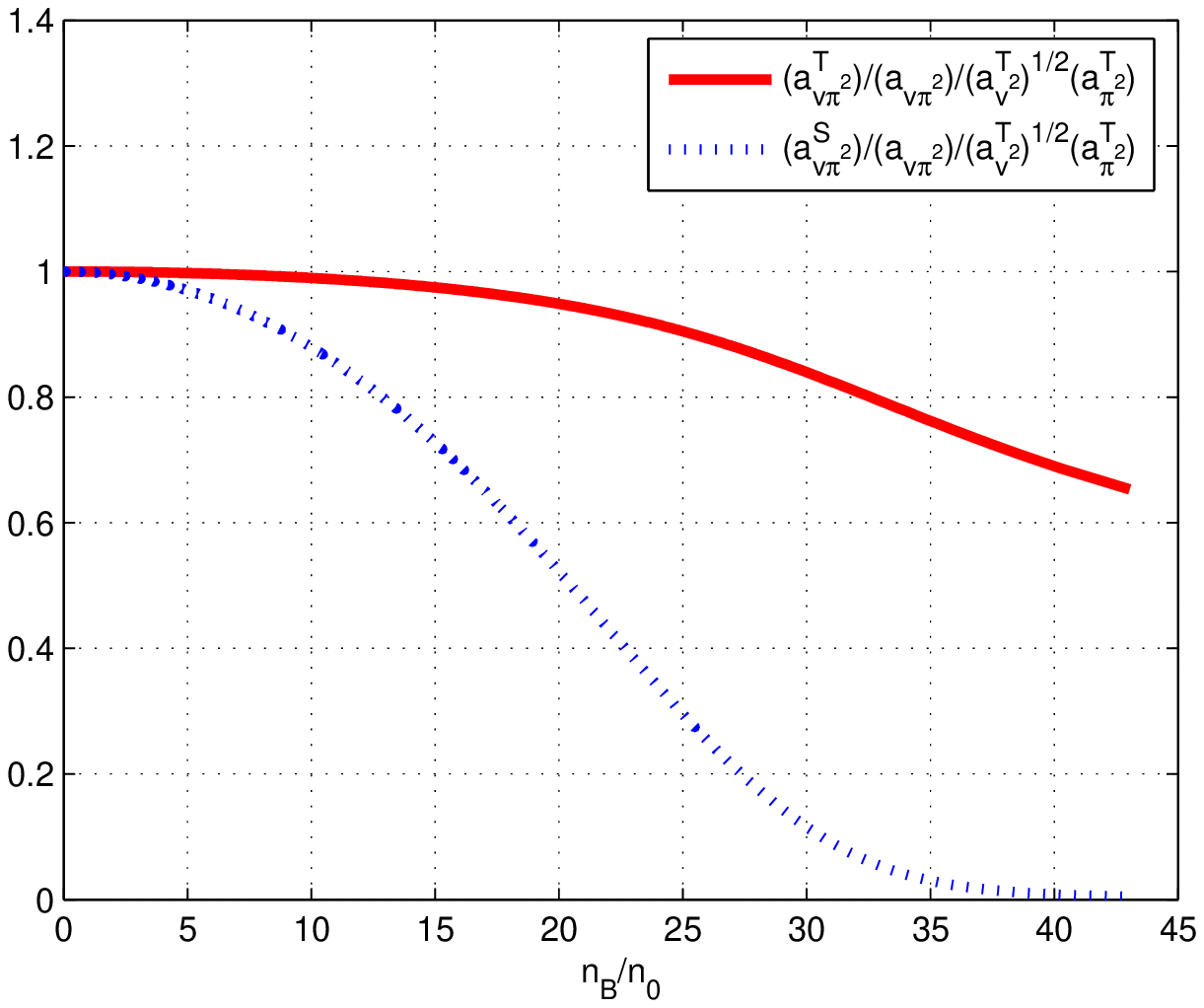}
   \includegraphics[width=7cm]{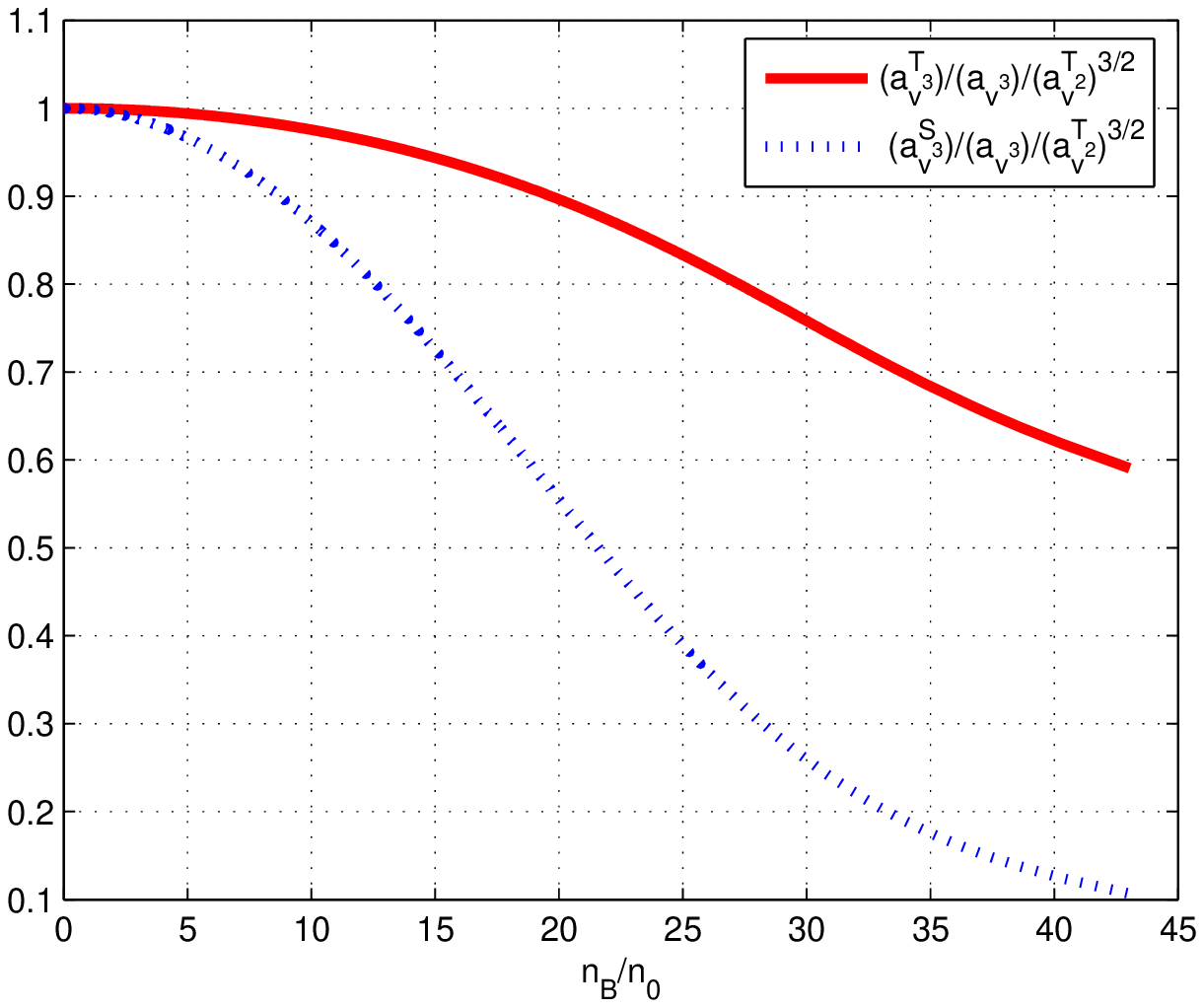}
  \caption{$V\pi\pi$ and $VVV$ couplings.}
  \label{fpi2tsvpp}
  \end{center}
\end{figure}

\begin{figure}[!hbp]
  \begin{center}
  \includegraphics[width=7cm]{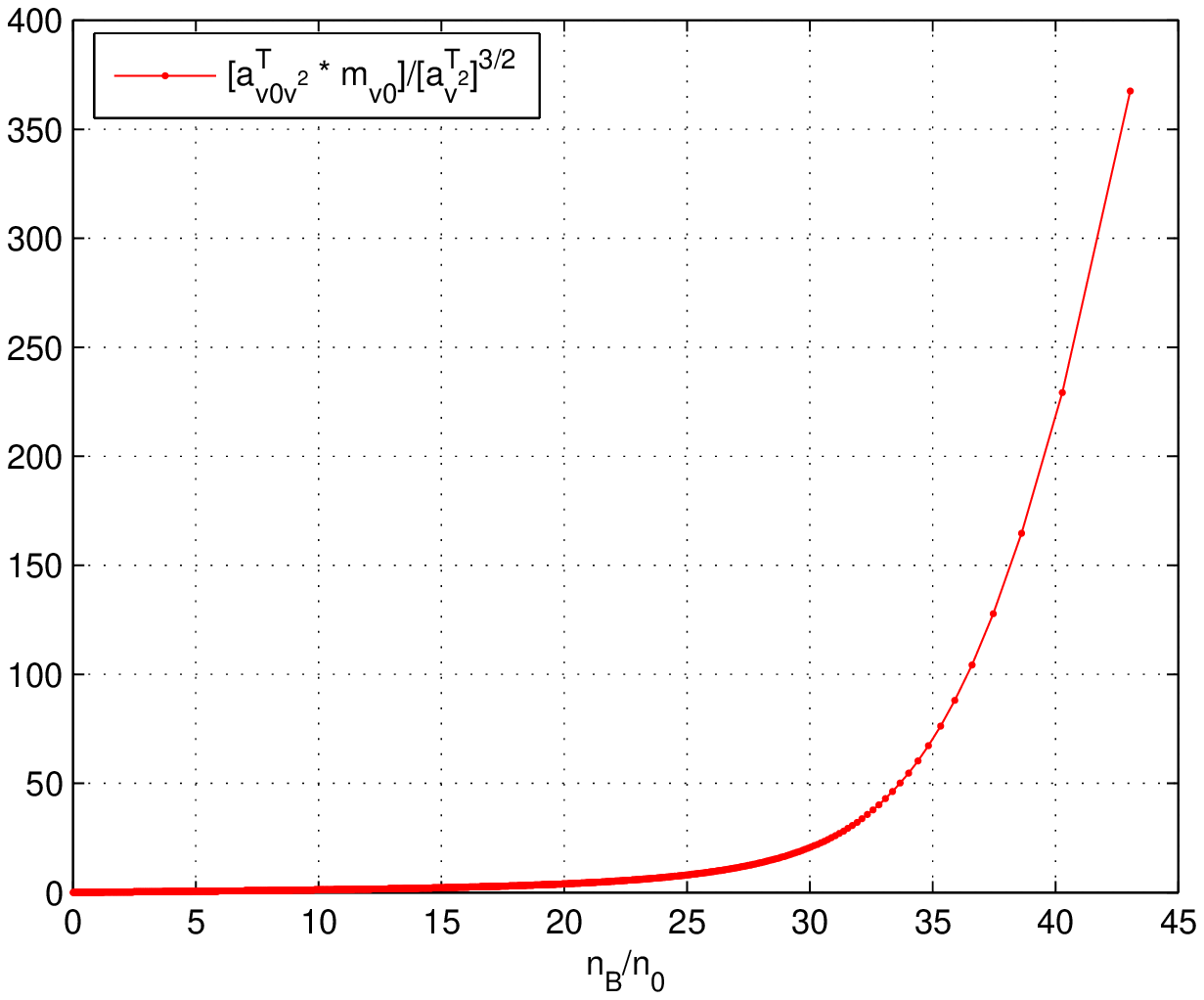}
  \includegraphics[width=7cm]{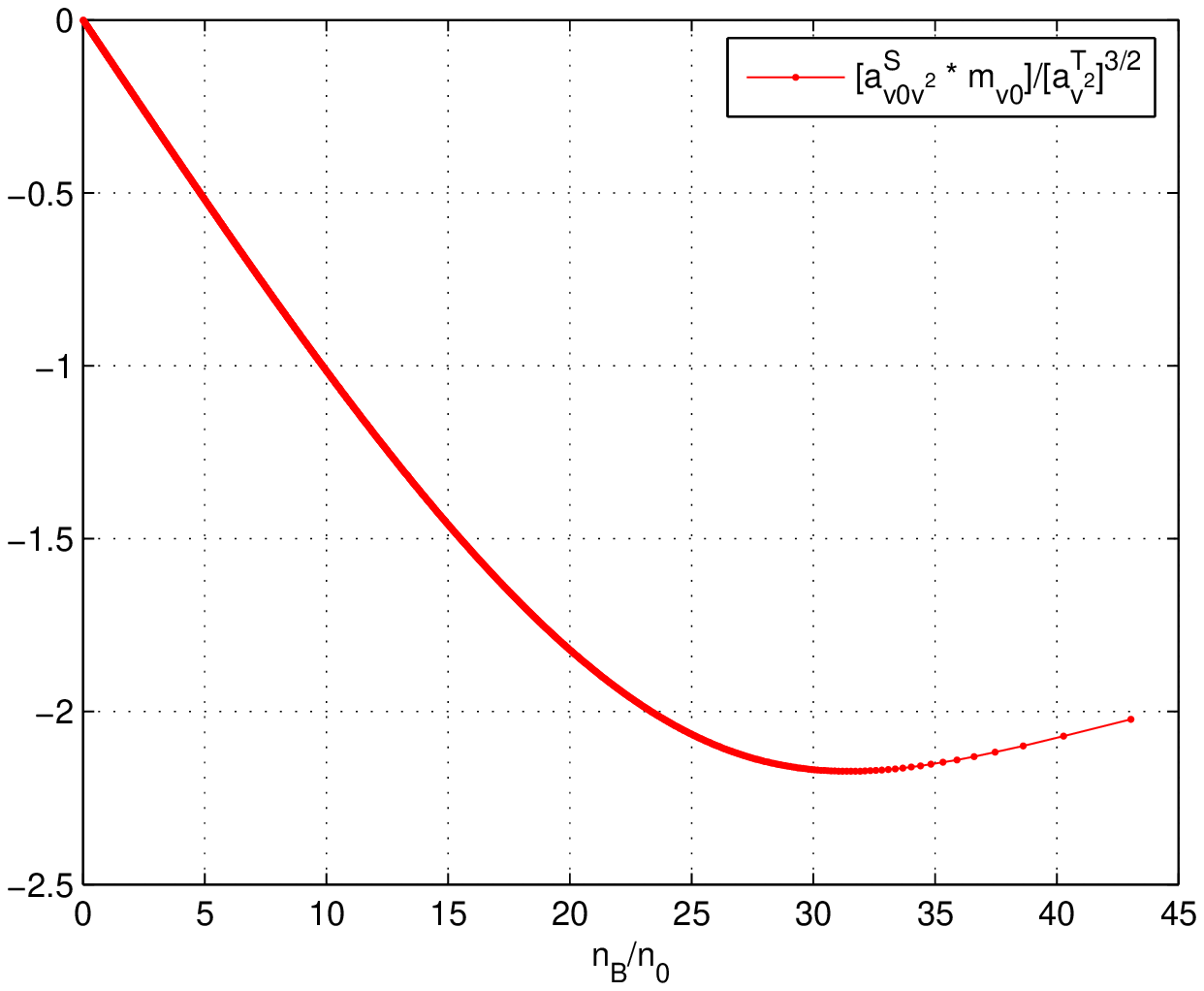}
  \includegraphics[width=7cm]{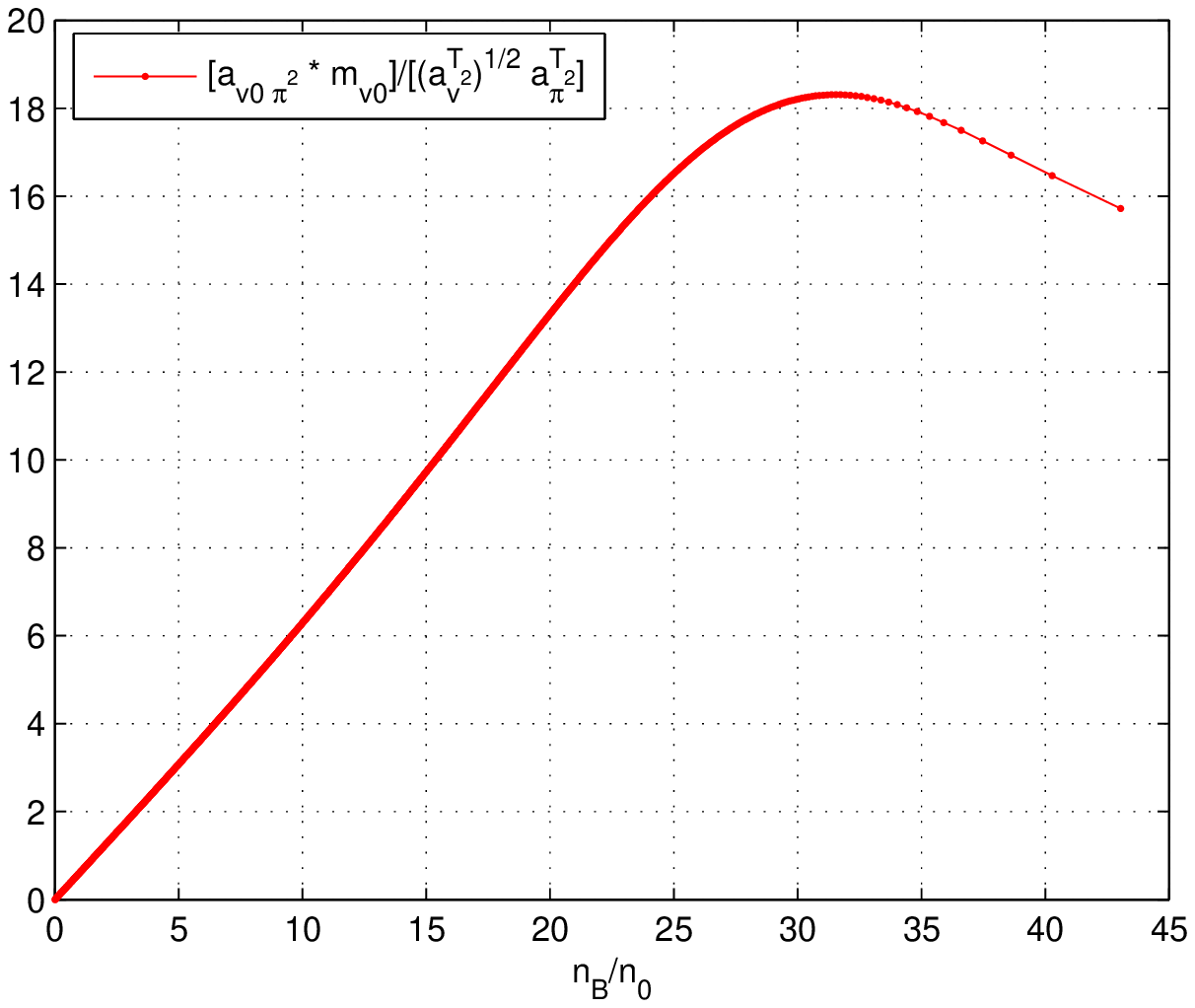}
  \includegraphics[width=7cm]{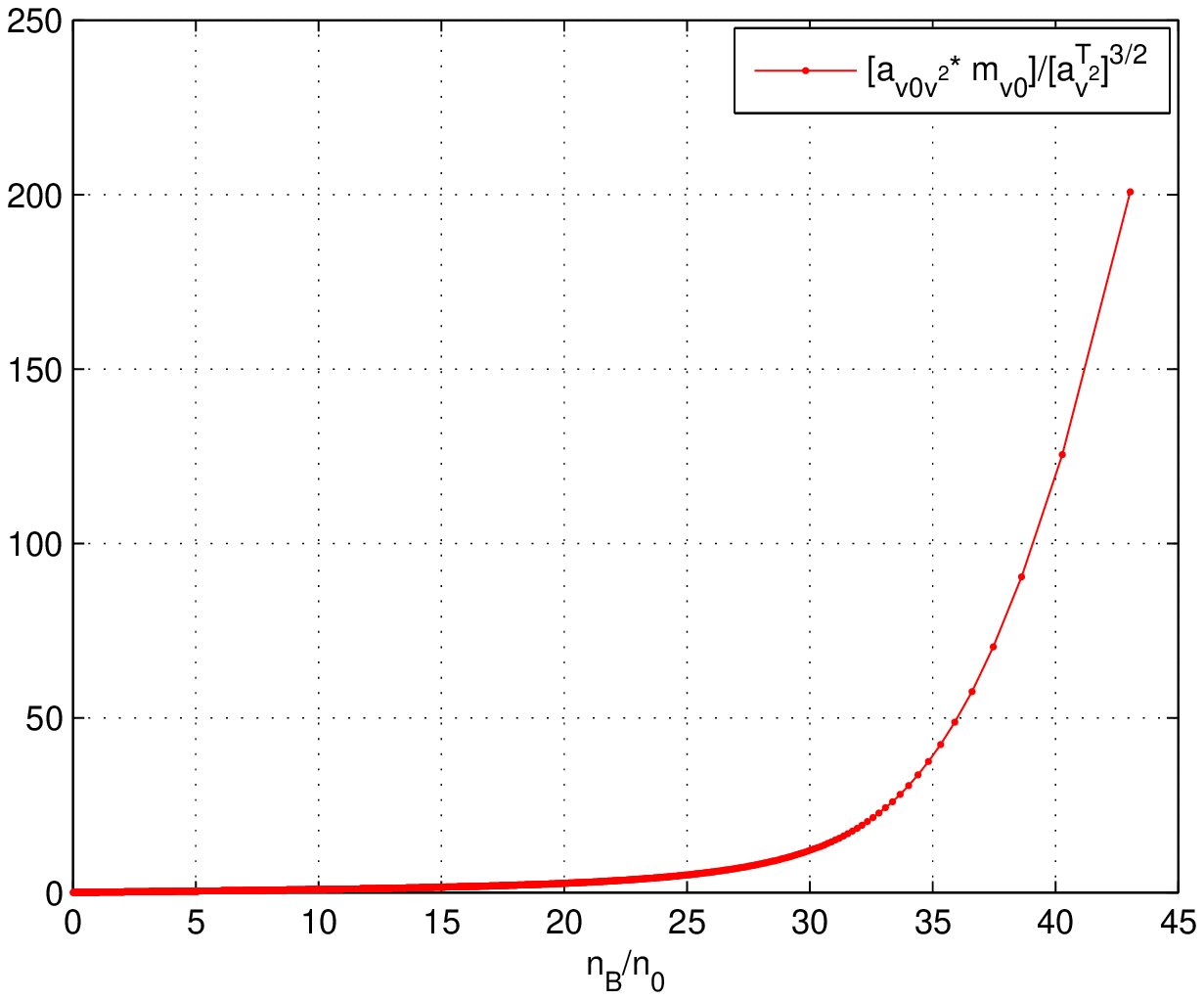}
  \caption{New terms}
  \label{fpi2tsnew}
  \end{center}
\end{figure}

The $V \pi\pi$ coupling is modified in matter. The longitudinal
and transverse couplings are shown in Fig.~\ref{fpi2tsvpp} (left).
The longitudinal coupling drops by 40\% at the freezing point,
while the transverse coupling {\it vanishes} when dense matter
freezes. In Fig.~\ref{fpi2tsvpp}  (right) $VVV$ coupling also decrease in matter.
The transverse couplings drop by 90\%, while the longitudinal
ones by 30\%. In addition to these standard couplings, new matter
dependent couplings emerge. Their composition and dependence
on the baryon density is shown in Fig.~\ref{fpi2tsnew}.

Many of the results presented here in matter bears similarities
(and differences) with arguments presented at finite temperature
using phenomenological models with hidden local symmetry~\cite{HIDDEN}.

\section{Conclusions}

We have provided a minimal extension of holographic
QCD~\cite{Sakai}
to dense matter. The induced DBI effective action on D8 in the
presence of a constant external $U(1)_v$ source ${\cal V}_0$ is
the effective action we expect from the gravity dual theory
from general principles. Indeed, if we were to solve for the effect
of the boundary ${\cal V}_0$ on the D4-brane background and the resulting
DBI action on the D8-probe brane, the answer is the {\it externally gauged}
DBI induced action in the vacuum. The effective action develops an
imaginary time at large ${\cal V}_0$, a signal that the D8-probe-brane
craks in the external field.

Holographic matter describes dense QCD at large $N_c$ with baryons
as solitons. Its bulk pressure asymptotes a constant at large density,
signalling total freezing with zero mean kinetic energy. Before freezing
the matter is dominated by two-body repulsion at low density and three-body
attraction at intermediate densities. The two-body effects are 100 times
stronger than the three-body effects.

In holographic matter the pions stall to almost a stop, while the
vector mesons only slow down. The vector masses drop by about 20\%,
while vector screening becomes increasingly large. The transverse
vector mesons completly decouple from pions at large densities. Many
of the current results bear similarities with known results from
effective models at large $N_c$. In a way, they are new as they
provide first principle calculations to large $N_c$ and strongly
coupled QCD.

In the work to follow, we will present results for holographic matter
including finite quark masses and temperature.

\section{Acknowledgements}

The work of KYK and IZ was supported in part by US-DOE grants
DE-FG02-88ER40388 and DE-FG03-97ER4014. The work of SJS was
supported by KOSEF Grant R01-2004-000-10520-0 and by SRC Program of
the KOSEF with grant number R11 - 2005- 021.

\end{document}